\documentclass[pra,twocolumn,amsmath,amssymb,superscriptaddress]{revtex4-1}
\usepackage{epsfig,amsmath}
\usepackage{subfigure}
\usepackage{graphicx}% Include figure files
\usepackage{dcolumn}% Align table columns on decimal point
\usepackage{stmaryrd}
\usepackage{mathrsfs}
\usepackage{pifont}
\usepackage{amsthm}
\usepackage{amssymb}
\usepackage{bm}
\usepackage{latexsym}
\usepackage[colorlinks=true,linkcolor=blue,citecolor=blue]{hyperref}
\usepackage{color}
\usepackage{epstopdf}

\begin{document}

\title{Accelerated adiabatic passage in cavity magnomechanics}

\author{Shi-fan Qi}
\affiliation{Department of Physics, Zhejiang University, Hangzhou 310027, Zhejiang, China}

\author{Jun Jing}
\email{jingjun@zju.edu.cn}
\affiliation{Department of Physics, Zhejiang University, Hangzhou 310027, Zhejiang, China}

\date{\today}

\begin{abstract}
Cavity magnomechanics provides a readily-controllable hybrid system, that consisted of cavity mode, magnon mode, and phonon mode, for quantum state manipulation. To implement a fast-and-robust state transfer between the hybrid photon-magnon mode and the phonon mode, we propose two accelerated adiabatic-passage protocols individually based on the counterdiabatic Hamiltonian for transitionless quantum driving and the Levis-Riesenfeld invariant for inverse engineering. Both the counterdiabatic Hamiltonian and the Levis-Riesenfeld invariant generally apply to the continuous-variable systems with arbitrary target states. It is interesting to find that our counterdiabatic Hamiltonian can be constructed in terms of the creation and annihilation operators rather than the system-eigenstates and their time-derivatives. Our protocol can be optimized with respect to the stability against the systematic errors of coupling strength and frequency detuning. It contributes to a quantum memory for photonic and magnonic quantum information. We also discuss the effects from dissipation and the counter-rotating interactions.
\end{abstract}

\date{\today}

\maketitle

\section{Introduction}\label{Introduction}

Hybrid cavity-magnon systems~\cite{cavitymagnonics,magnon,magnon2} base on the expedient control of coherent magnon-photon coupling have recently attracted intensive attention. They found new avenues for quantum computing~\cite{quantumcomputing}, quantum communication~\cite{quantumcommunication}, and quantum sensing~\cite{quantumsense}. Analog to the cavity quantum electrodynamics (QED)~\cite{circuit} and optomechanics~\cite{optcavity}, cavity magnomechanics~\cite{magnoncavity} develops rapidly to become a mesoscopic platform for quantum information processing in both theoretical~\cite{cavityyig1,cavityyig2,magnonblockade,quantumnetwork} and experimental aspects~\cite{yigcavity,yigcavity2,yigcavity3,yigcavity4,kerrmagnon,gatemagnon,magnonqubit,magnonqubit2}. Active investigations about magnon-based quantum information transfer focus on the coupling between photons and magnons and that between magnons and phonons in the ferrimagnetic material. Typical applications of these couplings include the hybrid entanglement and steering~\cite{mppentang,mppentang2,mppentang3,steermagnon}, the photon-phonon interface~\cite{intermagnon,adiabaticpmp}, and the magnomechanical phonon laser~\cite{phononlaser}.

In particular, a cavity magnomechanical system~\cite{magnoncavity} consists of a single-crystal yttrium iron garnet (YIG) sphere placed inside a microwave cavity, where the magnon modes formed by the excitations the collective angular momentum of the spins in such a magnetic-material sphere are coupled with the deformation phonon modes via a magnetostrictive force, and also with the electromagnetic cavity modes via a magnetic dipole interaction. The phonon in the YIG sphere decays with a rate about $100$ Hz~\cite{magnoncavity,mppentang}, much smaller than its own frequency and those of the magnon and the photon. That enables the storage-and-transfer of the microwave photonic and magnonic states as long-lasting modes, constituting a key step for the future quantum communication networks~\cite{storage1}. Inspired by the light-matter interface implemented within the cavity QED~\cite{statetransfer1,statetransfer2}, the optomechanical systems~\cite{statetransfer3,statetransfer4,statetransfer5}, and the optical waveguides~\cite{stawaveguide}, the stimulated Raman adiabatic passage between photon and phonon~\cite{adiabaticpmp} and the magnon-assisted photon-phonon conversion~\cite{intermagnon} have been proposed in the cavity magnomechanical systems. These protocols however demand a long evolution time and then the quantum system is prone to decoherence. Thus the shortcut-to-adiabatic (STA) protocols are desired to realize a quick-and-faithful state transfer in the cavity magnomechanical systems.

Various STA approaches~\cite{shortcut,STAtheory}, including the transitionless quantum driving (TQD) based on the counterdiabatic Hamiltonian~\cite{Berry2009,stacd,stacd2} and the inverse engineerings based on Lewis-Riesenfeld (LR) invariant~\cite{LRinvariant}, time-rescaling~\cite{timerescaling}, or noise-induced adiabaticity~\cite{noiseadia,noiseadia3,noiseadia2}, have been applied to several prototypes, such as, two- and three-level atomic system~\cite{staatom,staatom2,staatom3}, quantum harmonic oscillator~\cite{staoscillator}, optomechanical system~\cite{staoptomechanical,staopto,staopto2}, and coupled-waveguide device~\cite{stawaveguide}. Comparing to the existing STA methods, which are limited to the discrete systems or the continuous-variable systems in a subspace with a fixed excitation number~\cite{optimalcontrol,staopto,staopto2}, our protocol in this work is independent of the target state and can be applied to any coupled harmonic oscillators. The stability of our STA protocol for state transfer will be examined with respect to its robustness against the systematic errors~\cite{syserror}, that result mainly from the intensity fluctuations or inaccurate realization of the time-dependent driving laser. Then we can optimize the STA protocols in the cavity magnomechanical system~\cite{syserror,optimalcontrol,optimalcontro2,optimalcontro3}.

The rest part of this work is structured as following. In Sec.~\ref{secModel}, we introduce a hybrid quantum model for cavity magnomechanics and then provide the effective Hamiltonian describing the interaction between the hybrid photon-magnon mode and the phonon mode. The details of the derivation can be found in Appendix~\ref{appa}. Based on the effective Hamiltonian, we then propose two STA protocols in Sec.~\ref{STA} for fast quantum state transfer in the cavity magnomechanical systems. In Sec.~\ref{secTQD}, we construct the counterdiabatic Hamiltonian for TQD in terms of the creation and annihilation operators of the bosonic modes rather than the eigenstates of the effective Hamiltonian and their time-derivatives. In Sec.~\ref{secLR}, we derive a general Levis-Riesenfeld invariant for a two-coupled-bosonic-mode system, which applies to arbitrary target state. In Sec.~\ref{syserror}, we provide a general formalism for the systematic error of the effective Hamiltonian. The error sensitivity or robustness of various protocols, including the $\pi$-pulse, TQD and invariant-based STA, are analyzed and optimized in Secs.~\ref{pipulse}, \ref{Tqderror}, and \ref{lrerror}, respectively. In Sec.~\ref{discussion}, we discuss the effects of decoherence and counter-rotating interaction in the transport protocols by master equation and numerical simulation, respectively. The whole work is summarized in Sec.~\ref{conclusion}.

\section{Model}\label{secModel}

\begin{figure}[htbp]
\centering
\includegraphics[width=0.4\textwidth]{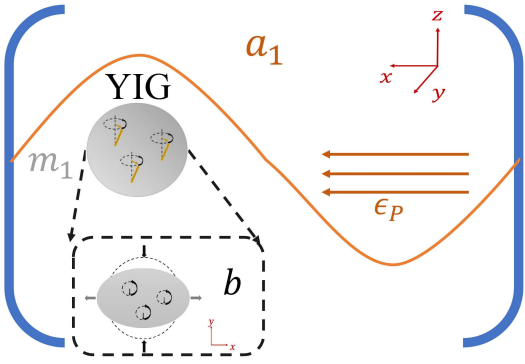}
\caption{Schematic diagram of a YIG sphere placed in a microwave cavity nearby the maximum magnetic field of the cavity mode. The uniform bias magnetic field exciting the Kittel mode in the YIG and establishing the magnon-photon coupling is aligned along the $z$-axis. The photon mode is driven by a microwave source along the $x$-axis (with a Rabi frequency $\epsilon_p$). The insects shows how the dynamic magnetization of magnon (vertical black arrows) causes the deformation (compression along the $y$ direction) of the YIG sphere (and vice versa), which rotates at the magnon frequency.}\label{Model}
\end{figure}

Consider a hybrid system in cavity-magnonic setup shown in Fig.~\ref{Model}, where a YIG sphere is inserted into a microwave cavity. The system is constituted by microwave-mode photons, magnons and mechancical-mode phonons, that has been experimentally realized in the dispersive regime. The magnons are coupled to photons via the Zeeman interaction and simultaneously coupled to phonons by the magnetization interaction. In particular, the temporally-varying magnetization induced by the magnon excitation inside the YIG sphere leads to the deformation of its geometrical structure, which forms the vibrational modes (phonons) of the sphere. The Hamiltonian of the full system is given by ($\hbar=1$)~\cite{magnoncavity}
\begin{equation}\label{Hmodel}
\begin{aligned}
H_0&=\omega_a a_1^\dagger a_1+\omega_m m_1^\dagger m_1+\omega_b b^\dagger b+g_{mb}m_1^\dagger m_1(b+b^\dagger)\\
&+g_{ma}(a_1m_1^\dagger+a_1^\dagger m_1)+i(\epsilon_p a_1^\dagger e^{-i\omega_p t}-\epsilon^*_p a_1e^{i\omega_p t}),
\end{aligned}
\end{equation}
where $a_1(a_1^\dagger)$, $m_1(m_1^\dagger)$ and $b(b^\dagger)$ are the annihilation (creation) operators of the microwave cavity mode, the magnon of the ground Kittel mode, and the mechanical mode with transition frequencies $\omega_a$, $\omega_m$ and $\omega_b$, respectively. The frequency of the magnon mode $\omega_m=\gamma h$, where $\gamma$ is the gyromagnetic ratio and $h$ is the external bias magnetic field. Thus the frequency $\omega_m$ can be readily tuned by the external magnetic field. $g_{ma}$ and $g_{mb}$ are respectively the single-excitation coupling strength of the photon-magnon interaction and magnon-phonon interaction. The last term in $H_0$ describes the external driving of the photon mode, where $\omega_p$ is the frequency of the driven and $\epsilon_p$ is the Rabi frequency of driving field.

Following the standard linearization approach~\cite{optcavity} and under the proper driving condition, we can extract an effective Hamiltonian describing the interaction between a hybridized photon-magnon mode and the phonon mode (the derivation detail can be found in Appendix~\ref{appa}):
\begin{equation}\label{Hlinear}
H=(\Delta-\omega_b)m^\dagger m+(g m^\dagger b+g^*mb^\dagger).
\end{equation}
Here $m=\sin\phi a_1-\cos\phi m_1$ is the hybridized normal mode with $\tan(2\phi)\equiv2g_{ma}/{(\omega_a-\omega_m)}$. $g$ is the driving-enhanced coupling strength between the hybrid mode-$m$ and the mechanical mode-$b$
\begin{equation}\label{Geffnormal}
g=g_{mb}m_{s}\cos^2\phi-g_{mb}a_{s}\sin\phi\cos\phi,
\end{equation}
where
\begin{equation}\label{steadynormals}
m_s=\frac{\epsilon_p\sin\phi}{i\Delta+\kappa_m}, \quad
a_s=\frac{\epsilon_p\cos\phi}{i\Delta'+\kappa_a},
\end{equation}
with the effective frequencies of the hybridized modes
\begin{equation}
\begin{aligned}
\Delta=\frac{\omega_a+\omega_m}{2}-\omega_p-\sqrt{\left(\frac{\omega_a-\omega_m}{2}\right)^2+g^2_{ma}},\\
\Delta'=\frac{\omega_a+\omega_m}{2}-\omega_p+\sqrt{\left(\frac{\omega_a-\omega_m}{2}\right)^2+g^2_{ma}},\\
\end{aligned}
\end{equation}
and the decay rates $\kappa_m$ and $\kappa_a$.

The coupling $g$ can then be varied by tuning the driving parameters $\epsilon_p$ and $\omega_p$. The hybrid-mode frequency can be altered by adjusting the strength of the external magnetic bias field~\cite{mppentang}. Thus both terms in Eq.~(\ref{Hlinear}) can be modulated with time. The following state transfer between the hybrid mode and the mechanical mode could be formally started from the time-dependent Hamiltonian
\begin{equation}\label{Hamiltonian0}
H(t)=\Delta(t)m^{\dagger}m+g(t)m^\dagger b+g^*(t)b^\dagger m.
\end{equation}
where $\Delta(t)\equiv\Delta-\omega_b$. In the framework of various STA state-transfer protocols, it is instructive to transform the system Hamiltonian $H(t)$ into the rotating frame with respect to $U=\exp[i\int_0^tds\Delta(s)/2(m^\dagger m-b^\dagger b)]$,
\begin{equation}\label{Hamiltonian}
H(t)=\frac{\Delta(t)}{2}\left(m^{\dagger}m-b^{\dagger}b\right)+g(t)m^\dagger b+g^*(t)b^\dagger m.
\end{equation}

\section{The state transfer protocols}\label{STA}

A straightforward protocol to achieve the state transfer is using a $\pi$ pulse. In this case, one can hold the driving frequency to be resonant with the hybrid mode $m$, i.e., $\Delta(t)=0$ for all of time in Hamiltonian~(\ref{Hamiltonian}). Then a initial state $|\psi(0)\rangle=(\sum_kC_k|k\rangle_m)|0\rangle_b$ with arbitrary normalized coefficients $C_k$'s could be converted to
\begin{equation}\label{pitarget}
|\psi(T)\rangle=|0\rangle_m\left(\sum_kC_k e^{-i\frac{k\pi}{2}}|k\rangle_b\right)
\end{equation}
after a desired period $T$ as long as the coupling strength satisfies $\int^T_0 dt|g(t)|=\pi/2$. For example, for a flat $\pi$ pulse we can set $g(t)=\pi/(2T)$. Note the final state for the mechanical mode in $|\psi(T)\rangle$ is not exactly the same one for the hybrid mode $m$ in the initial state $|\psi(0)\rangle$, regarding the dynamical phase $k\pi/2$. However, the phase difference between the final and the initial states could be compensated by the dynamical local phase $e^{-ik\omega_b\tau}$ after a free evolution time $\tau$ by $H_b=\omega_b b^\dagger b$. The phase difference vanished at the moments satisfying $\omega_b\tau+\pi/2=2n\pi$ with integer $n$. We therefore do not distinguish the distinction between the states $\sum_kC_ke^{-i\frac{k\pi}{2}}|k\rangle$ and $\sum_kC_k|k\rangle$ when calculating the state-transfer fidelity. The state-transfer fidelity or efficiency is thus measured by the target-state population
\begin{equation}\label{population}
P=\sum_{C_k\neq0}|\langle 0k|\psi(t)\rangle|^2,
\end{equation}
where $|\psi(t)\rangle$ is the dynamical state determined by the initial state $|\psi(0)\rangle$ and the Hamiltonian~(\ref{Hamiltonian}). Note when the target state is a Fock state $|N\rangle$, $P$ becomes the conventional fidelity.

The $\pi$-pulse protocol is straightforward but sensitive to both decoherence and the systematic errors~\cite{syserror} caused by the long-time evolution and the fluctuations of the Hamiltonian, respectively. The state-transfer protocol can by improved by the accelerated adiabatic passage or shortcuts to adiabaticity that is robust to both decoherence and systematic error. In the two subsequent subsections, we will introduce two STA protocols about TQD and LR-invariant, to our hybrid magnomechanical model. we apply our results to two extreme examples of the quantum state of mode-$m$ and mode-$b$. One is based on the number state or Fock state and the other is on the cat state as both superposed coherent state and Fock state.

\subsection{Transitionless quantum driving for continuous-variable system}\label{secTQD}

The TQD approach was proposed in the first decade of this century, depending on the full knowledge about the instantaneous eigenstructure of the original Hamiltonian. Conventionally, if the original time-dependent Hamiltonian $H(t)$ could be formally expressed in the spectral representation as $H(t)=\sum_{n}E_n(t)|n(t)\rangle\langle n(t)|$, then assisted by an ancillary Hamiltonian, or called the counterdiabatic (CD) Hamiltonian~\cite{Berry2009,shortcut}
\begin{eqnarray}\label{CD}
H_{\rm CD}(t)=i\sum_n\left[1-|n(t)\rangle\langle n(t)|\right]|\dot{n}(t)\rangle\langle n(t)|,
\end{eqnarray}
the system could keep track of the instantaneous eigenstates of $H(t)$ in a much faster speed. This approach is also believed to highly robust against the control-parameter variations~\cite{shortcut}. One can understand that it applies usually to the discrete systems, since it is explicitly represented by the eigenstates and their time-derivatives.

In this work, we derive the CD term using operators in the continuous-variable systems. Using the Bogolyubov transformation~\cite{bogo}, the Hamiltonian in Eq.~(\ref{Hamiltonian}) is rewritten as
\begin{equation}\label{ham}
H(t)=\omega_A A^\dagger A+\omega_B B^\dagger B,
\end{equation}
where $\omega_{A,B}=\pm\sqrt{\Delta^2+4g^2}/2$ and
\begin{equation}\label{AB}
\begin{aligned}
&A\equiv\cos\theta m+\sin\theta b, \\ &B\equiv\sin\theta m-\cos\theta b
\end{aligned}
\end{equation}
with $\tan(2\theta)=2g(t)/\Delta(t)$. To simplify the formation of the TQD protocol, the coupling strength $g(t)$ in this protocol is set to be real, i.e., $g(t)=g^*(t)$.

In the subspace with a fixed and arbitrary excitation number $N$, the CD Hamiltonian for the system Hamiltonian in Eq.~(\ref{ham}) can be written as $H_{\rm CD}=i\sum_{n=0}^N|\dot{\epsilon}_n\rangle\langle\epsilon_n|$, where the orthonormal eigenstates read
\begin{equation}\label{En}
|\epsilon_{N-n}\rangle=\frac{1}{\sqrt{(N-n)!n!}}(A^\dagger)^{N-n}(B^\dagger)^n|0\rangle,
\end{equation}
with $|0\rangle$ the vacuum state for both modes. Due to the fact that
\begin{equation}\label{ABdot}
\begin{aligned}
&\dot{A}=-\dot{\theta}\sin\theta m+\dot{\theta}\cos\theta b=-\dot{\theta}B, \\
&\dot{B}=\dot{\theta}\cos\theta m+\dot{\theta}\sin\theta b=\dot{\theta}A,
\end{aligned}
\end{equation}
we have
\begin{equation}\label{enkdot}
\begin{aligned}
|\dot{\epsilon}_{N-n}\rangle&=-\dot{\theta}\sqrt{(n+1)(N-n)}|\epsilon_{N-n-1}\rangle\\
&+\dot{\theta}\sqrt{n(N-n+1)}|\epsilon_{N-n+1}\rangle.
\end{aligned}
\end{equation}
Then the CD Hamiltonian by Eq.~(\ref{CD}) becomes
\begin{equation}\label{Hcds}
\begin{aligned}
H_{\rm CD}&=i\sum_{n=0}^N|\dot{\epsilon}_n\rangle\langle\epsilon_n|\\
&=-\dot{\theta}\sum_{n=0}^{N-1}\sqrt{(N-n)(n+1)}|\epsilon_{N-n-1}\rangle\langle\epsilon_{N-n}|\\
&+\dot{\theta}\sum_{n=1}^N\sqrt{n(N-n+1)}|\epsilon_{N-n+1}\rangle\langle\epsilon_{N-n}|.
\end{aligned}
\end{equation}
According to the definition in Eq.~(\ref{En}), the first term in Eq.~(\ref{Hcds}) expands
\begin{equation}\label{sumeigen}
\begin{aligned}
&\sum_{n=0}^{N-1}\sqrt{(N-n)(n+1)}|\epsilon_{N-n-1}\rangle\langle \epsilon_{N-n}|\\
=&\sum_{n=0}^{N-1}\sqrt{(N-n)(n+1)}\frac{(A^\dagger)^{N-n-1}(B^\dagger)^{n+1}A^{N-n}B^n}{(N-n-1)!n!\sqrt{(N-n)(n+1)}}\\
=&\sum_{n=0}^{N-1}\frac{(A^\dagger A)^{N-n-1}(B^\dagger B)^{n}B^\dagger A}{(N-n-1)!n!},\\
=&\frac{(A^\dagger A+B^\dagger B)^{N-1}}{(N-1)!} B^\dagger A=IB^\dagger A=B^\dagger A.
\end{aligned}
\end{equation}
where $I$ is the identify operator in the subspace with $N-1$ excitations. In the last line, we have applied the binomial theorem $(x+y)^N=\sum^{N}_{n=0}C^n_N x^ny^{N-n}, C_N^n=N!/[n!(N-n)!]$. Similarly, the second term in Eq.~(\ref{Hcds}) turns out to be
\begin{equation}\label{sumeigen2}
\sum_{n=1}^N\sqrt{n(N-n+1)}|\epsilon_{N-n+1}\rangle\langle\epsilon_{N-n}|=A^\dagger B.
\end{equation}

Note $N$ is arbitrary, then in the whole Hilbert space, the CD Hamiltonian could be expressed as
\begin{equation}\label{Hcdss}
H_{\rm CD}=i\dot{\theta}(A^\dagger B-B^\dagger A)=i\dot{\theta}(b^\dagger m-m^\dagger b).
\end{equation}
And regarding the original Hamiltonian (\ref{Hamiltonian}), then the total Hamiltonian for transitionless quantum driving reads
\begin{equation}\label{Htotalcd}
\begin{aligned}
&H_{\rm tot}=H(t)+H_{\rm CD} \\
&=\frac{\Delta(t)}{2}(m^\dagger m-b^\dagger b)+[g(t)-i\dot{\theta}]m^\dagger b+[g(t)+i\dot{\theta})]mb^\dagger,
\end{aligned}
\end{equation}
where the time-dependence and boundary-condition of
\begin{equation}\label{thetadot}
\dot{\theta}=\frac{\dot{g}\Delta-\dot{\Delta}g}{\Delta^2+4g^2}
\end{equation}
determines the speed of the accelerated adiabatic passage.

By virtue of the definitions in Eqs.~(\ref{AB}) and (\ref{En}), the adiabatic path from $|k\rangle_m|0\rangle_b$ to $|0\rangle_m|k\rangle_b$ under the total Hamiltonian~(\ref{Htotalcd}) is constructed by
\begin{equation}\label{kteigen}
|k(t)\rangle=|\epsilon_k(t)\rangle=\frac{1}{\sqrt{k!}}(A^\dagger)^k|0\rangle=|\epsilon_k\rangle
\end{equation}
under the boundary condition $\theta(t=0)=0$ and $\theta(t=T)=\pi/2$. And the system wavefunction can be written as
\begin{equation}\label{psikt}
|\psi_k(t)\rangle=e^{-i\chi_k(t)}|k(t)\rangle,
\end{equation}
where the quantum phase is
\begin{equation}\label{chiphase}
\begin{aligned}
\chi_k(t)&=\int^t_0dt'E_k(t')-i\int^t_0 dt'\langle\epsilon_k(t')|\partial_{t'}\epsilon_k(t')\rangle \\
&=\int^t_0 dt'E_k(t')=k\chi(t).
\end{aligned}
\end{equation}
Note $E_k(t)$ is the instantaneous eigenvalue of $|\epsilon_k\rangle$, $\langle\epsilon_k(t')|\partial_{t'}\epsilon_k(t')\rangle=0$, and $\chi(t)\equiv\int^t_0 dt'\omega_A(t')$. Alternatively, by setting $\theta(0)=\pi/2$ and $\theta(T)=0$, the adiabatic path followed by the system can be constructed by $|k(t)\rangle=|\epsilon_0(t)\rangle=\frac{1}{\sqrt{k!}}(B^\dagger)^k|0\rangle$. To be self-consistent, we stick to the boundary condition $\theta(0)=0$ and $\theta(T)=\pi/2$ in this work.

In general situations, a superposed state $|\psi(0)\rangle=\sum_kC_k|k0\rangle$ with normalized coefficients $C_k$'s will adiabatically evolve to
\begin{equation}\label{psit}
|\psi(T)\rangle=\sum_kC_ke^{-ik\chi(T)}|0k\rangle,
\end{equation}
at the desired moment $T$ and $C_k$ is invariant with time. The quantum phase $\chi_k(T)$ for the Fock state with $k$ excitations is proportional to $k$. By the preceding analyse, the phase difference can be periodically cancelled by the bare Hamiltonian of the mechanical mode. The state transfer thus has been indeed completed by Eq.~(\ref{psit}).

\begin{figure}[htbp]
\centering
\includegraphics[width=0.22\textwidth]{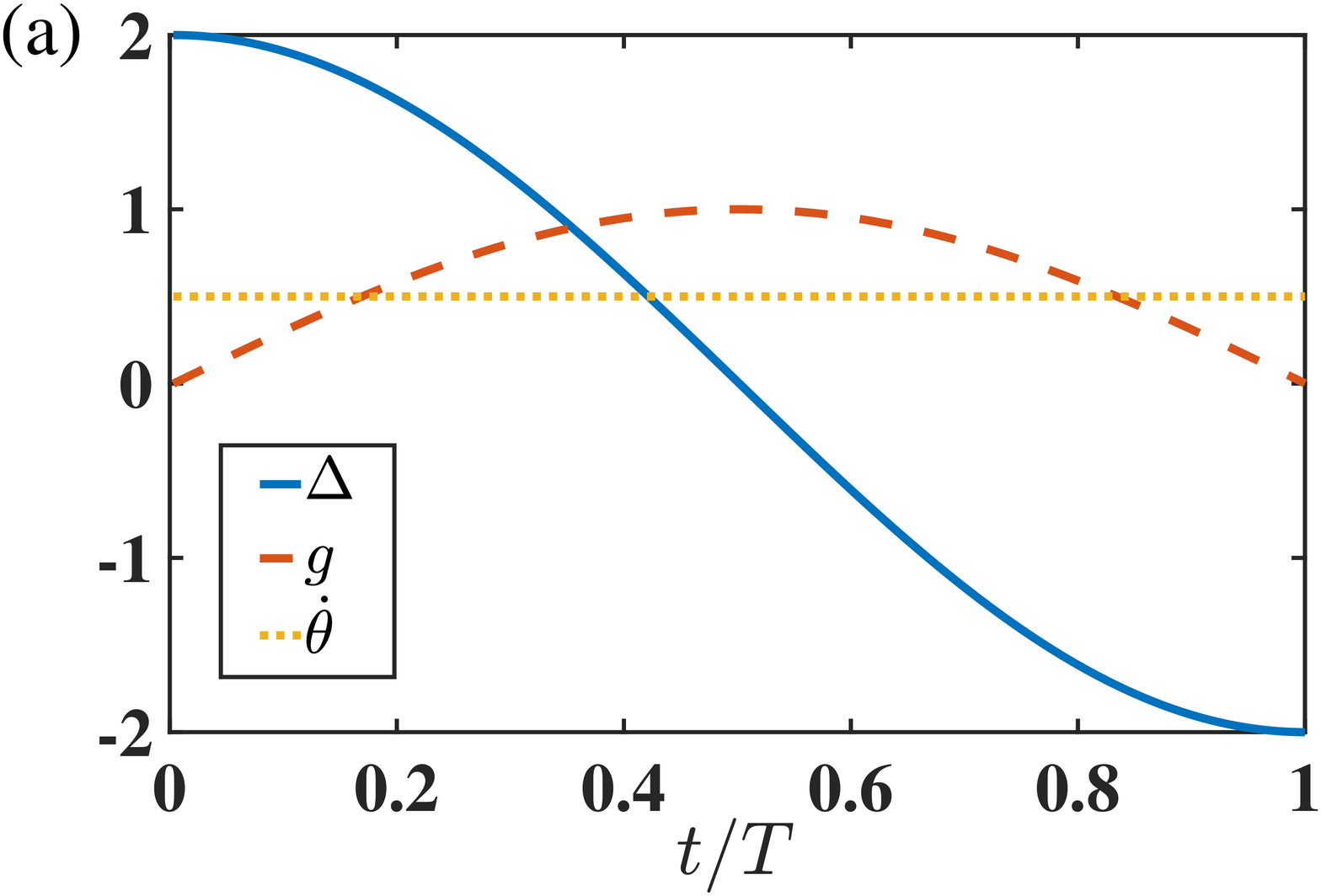}
\includegraphics[width=0.22\textwidth]{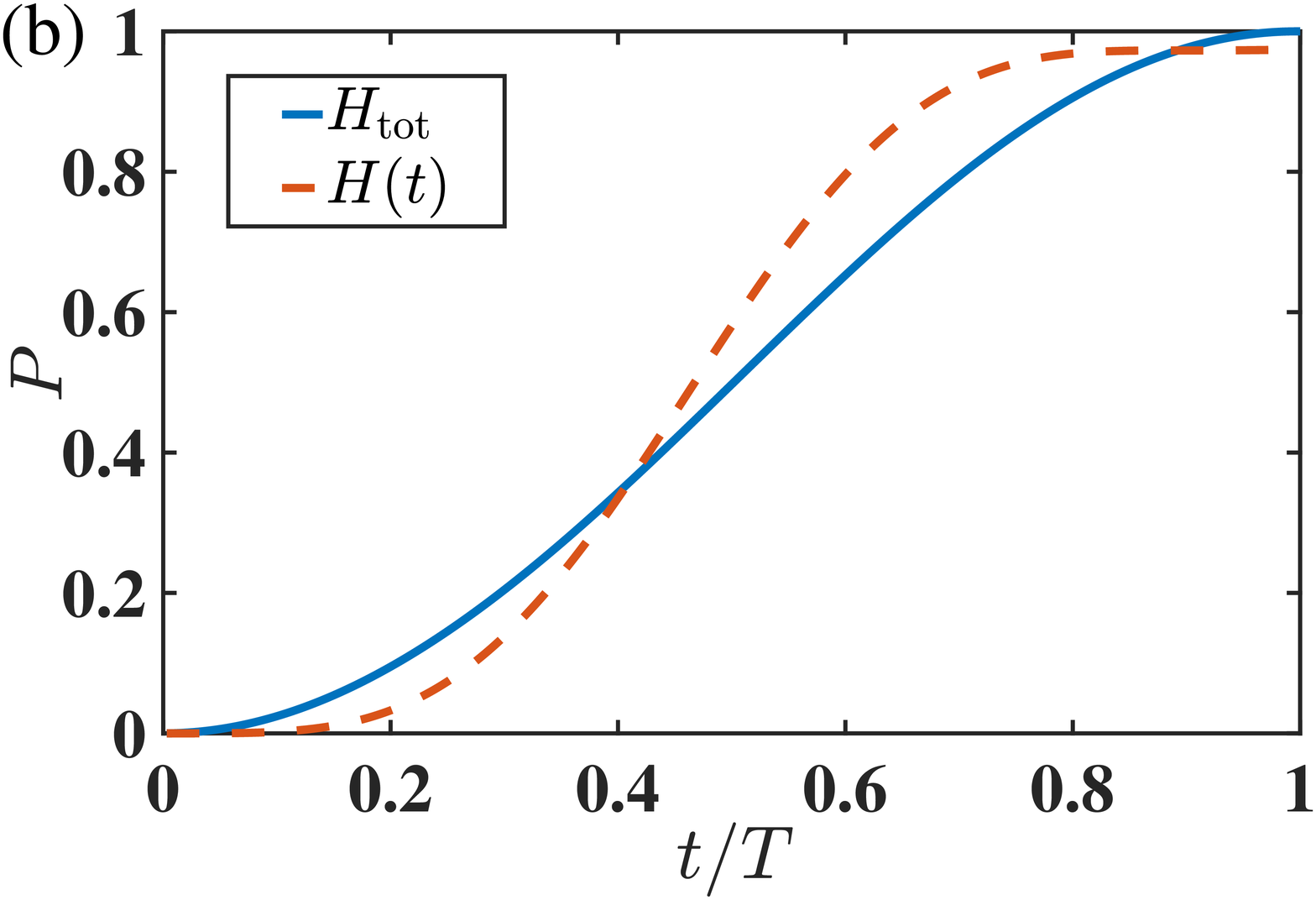}
\includegraphics[width=0.22\textwidth]{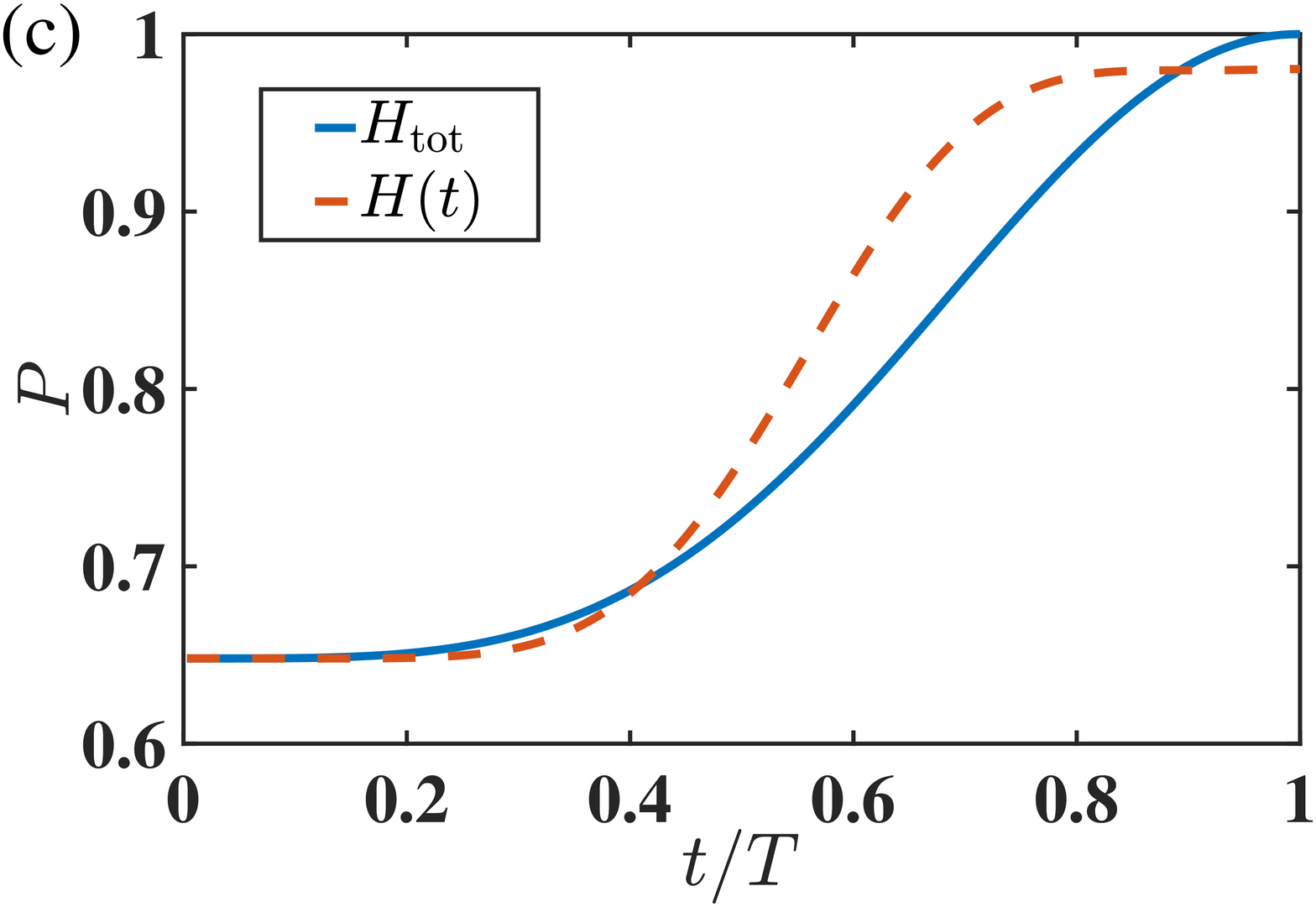}
\includegraphics[width=0.22\textwidth]{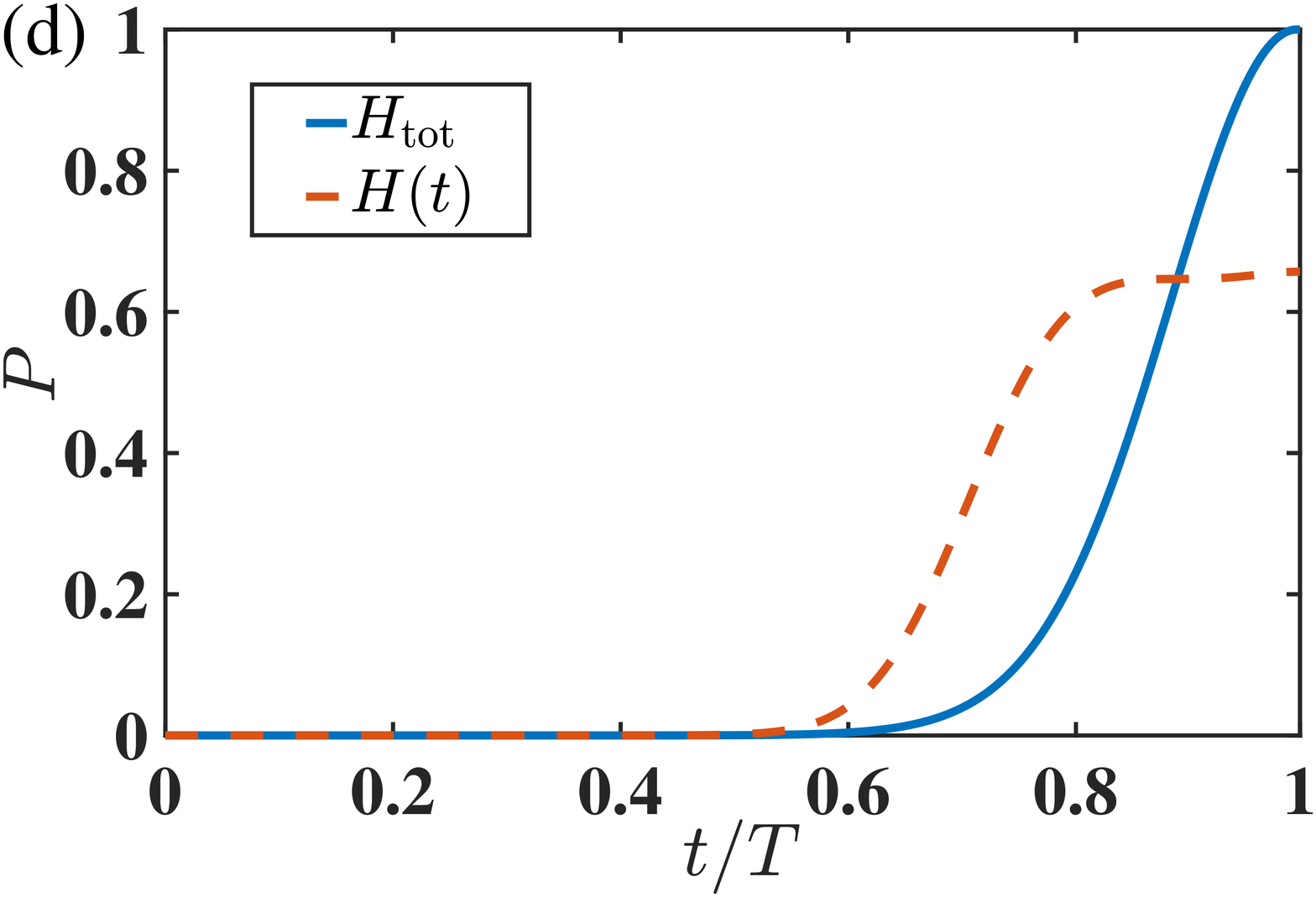}
\caption{(a): The time dependence of the driving-enhanced coupling strength $g$, the effective frequency of the lower-frequency hybrid mode $\Delta$, and the time-derivative of the control-parameter $\dot{\theta}$, in units of a coupling strength $\Omega=\pi/T$. (b), (c), and (d): The target-state population of the phonon mode $b$ for the initial state of the hybrid mode $m$ prepared as the Fock-state $|1\rangle$, the cat state with $\zeta=1$, and the cat state with $\zeta=4$, respectively.}\label{onecd}
\end{figure}

Now we can verify the TQD approach in the state transfer from mode-$m$ to mode-$b$ by presenting the practical dynamics of the target-state population $P$ given by Eq.~(\ref{population}). We choose the time-dependence of the effective frequency $\Delta$ and the driving-enhanced coupling strength $g$ to be in a sinusoid shape, and by Eq.~(\ref{thetadot}),
\begin{equation}\label{gGDelta1}
\Delta=2\Omega\cos(2\theta), \quad g=\Omega\sin(2\theta), \quad \theta=\frac{\pi}{2}\frac{t}{T},
\end{equation}
where $\Omega$ is the coupling strength determined by the desired transfer time $T$ as $\Omega T=\pi$. The shape functions of $\Delta$, $g$, and $\dot{\theta}$ are plotted in Fig.~\ref{onecd}(a). With various initial states of mode-$m$, the blue solid lines and the red dashed lines in Figs.~\ref{onecd}(b), (c), and (d) describe the dynamics of the target-state population under $H_{\rm tot}$ in Eq.~(\ref{Htotalcd}) with the counterdiabatic term and that under $H(t)$ in Eq.~(\ref{Hamiltonian}) without the counterdiabatic term, respectively.

One can find that practically the accelerated state-transfer could be perfectly completed by the TQD approach for either Fock-state and superposed states. The latter of the hybrid mode is prepared as an even cat-state $(|\zeta\rangle+|-\zeta\rangle)/\sqrt{2+2e^{-2\zeta^2}}$, where $|\zeta\rangle$ is the Glauber coherent state. For the Fock-state transfer $|1\rangle_m|0\rangle_b\rightarrow|0\rangle_m|1\rangle_b$ in Fig.~\ref{onecd}(b), $P$ approaches $0.97$ by the original Hamiltonian $H(t)$. In Figs.~\ref{onecd}(c) and (d) for the cat state with $\zeta=1$ and $4$, respectively, it is found that the TQD approach manifests its power for larger cat-state by achieving unit transfer population. In contrast, under the original Hamiltonian $H(t)$, $P$ approaches respectively $0.98$ and $0.65$ for $\zeta=1$ and $4$.

\subsection{Invariant-based inverse engineering}\label{secLR}

Another main-stream accelerated adiabatic-passage is the invariant-based inverse engineering~~\cite{shortcut,LRinvariant}, where the parametrical adiabatic-path of the system is designed through a Hermitian operator $I(t)$ termed the Levis-Riesenfeld invariant. For an arbitrary original Hamiltonian $H(t)$, the invariant satisfies
\begin{equation}\label{LRinvariant}
\frac{\partial I(t)}{\partial t}=-i[H(t), I(t)].
\end{equation}
In the framework of invariant-based inverse engineering, the wavefunction of a time-dependent Schr\"odinger equation $i\partial_t|\psi(t)\rangle=H(t)|\psi(t)\rangle$ can be expressed as $|\psi(t)\rangle=\sum_n C_ne^{-i\kappa_n(t)}|\epsilon_n(t)\rangle$, where $C_n$ is a time-independent amplitude, $|\epsilon_n(t)\rangle$ is the eigenstates of the invariant $I(t)$, and $\kappa_n$ is the Lewis-Riesenfeld phase defined by
\begin{equation}\label{kappa}
\dot{\kappa}_n(t)=\langle\epsilon_n(t)|[-i\partial_t+H(t)]|\epsilon_n(t)\rangle.
\end{equation}
To ensure the desired state transfer rather than the full-time adiabatic passage, $I(t)$ and $H(t)$ have to share the same eigenstates at both initial and final moments.

To carry out the derivation of a general LR invariant for our two-coupled-resonator system with arbitrary target state, we rewrite the system Hamiltonian $H(t)$~(\ref{Hamiltonian}) into
\begin{equation}\label{HgG}
\begin{aligned}
H(t)&=\frac{\Delta(t)}{2}m^\dagger m+[g_R(t)-ig_I(t)]m^\dagger b\\
&+[g_R(t)+ig_I(t)]mb^\dagger-\frac{\Delta(t)}{2}b^\dagger b.
\end{aligned}
\end{equation}
where $g_R(t)$ and $g_I(t)$ represent the real and imaginary parts of the complex coupling strength $g(t)$.

Its corresponding Lewis-Riesenfeld invariant can be formulated by
\begin{equation}\label{Lr}
\begin{aligned}
I(t)&=\cos\beta(m^\dagger m-b^\dagger b)+\sin\beta(e^{-i\alpha} m^\dagger b+e^{i\alpha}mb^\dagger), \\
&=A^\dagger A-B^\dagger B,
\end{aligned}
\end{equation}
where
\begin{equation}\label{operatorABLr}
\begin{aligned}
A=\cos\left(\frac{\beta}{2}\right)e^{\frac{i\alpha}{2}}m+\sin\left(\frac{\beta}{2}\right)e^{-\frac{i\alpha}{2}}b,\\
B=\sin\left(\frac{\beta}{2}\right)e^{\frac{i\alpha}{2}}m-\cos\left(\frac{\beta}{2}\right)e^{-\frac{i\alpha}{2}}b,\\
\end{aligned}
\end{equation}
are the normalized annihilation operators for $I(t)$ and both $\beta\equiv\beta(t)$ and $\alpha\equiv\alpha(t)$ are time-dependent functions to be determined. Substituting Eq.~(\ref{Lr}) into Eq.~(\ref{LRinvariant}), we have
\begin{equation}\label{thetaalpha}
\begin{aligned}
\dot{\beta}&=2g_I\cos\alpha-2g_R\sin\alpha,\\
\dot{\alpha}&=\Delta-\cot\beta(2g_R\cos\alpha+2g_I\sin\alpha).
\end{aligned}
\end{equation}
With the annihilation and creation operators of $I(t)$, the original Hamiltonian (\ref{HgG}) can be rewritten as
\begin{equation}\label{HAB}
H=\omega(A^\dagger A-B^\dagger B)+g_{AB}A^\dagger B+g^*_{AB}B^\dagger A,
\end{equation}
where
\begin{equation}
\begin{aligned}
\omega&\equiv\frac{\Delta\cos\beta}{2}+g_R\sin\beta\cos\alpha+g_I\sin\beta\sin\alpha,\\
g_{AB}&\equiv\frac{\Delta\sin\beta}{2}-g_R(\cos\beta\cos\alpha+i\sin\alpha) \\
&-g_I(\cos\beta\sin\alpha-i\cos\alpha).
\end{aligned}
\end{equation}
And then in the subspace with a fixed excitation number $N$, the general solution of the Schr\"odinger equation can be expressed as a superposition of the eigenstates of the invariant,
\begin{equation}\label{phi}
|\psi_N(t)\rangle=\sum^N_{n=0}p_n|\epsilon_{N-n}\rangle e^{-i\kappa_{N-n}(t)},
\end{equation}
where $p_n$ is a normalized coefficient, $|\epsilon_{N-n}\rangle$ is the normalized eigentstates of the LR invariant taking the same form as in Eq.~(\ref{En}), and $\kappa_{N-n}(t)$ is the quantum phase defined in Eq.~(\ref{kappa}).

We consider the state transfer from $|N\rangle_m|0\rangle_b$ to $|0\rangle_m|N\rangle_b$. With the definitions in Eqs.~(\ref{operatorABLr}) and (\ref{En}), one can construct a particular solution via
\begin{equation}\label{phit}
|\psi_N(t)\rangle=|\epsilon_N\rangle e^{-i\kappa_N(t)}=\frac{e^{-i\kappa_N(t)}}{\sqrt{N!}}(A^\dagger)^N|0\rangle.
\end{equation}
under the boundary conditions $\beta(0)=0$ and $\beta(T)=\pi$. As for the Lewis-Riesenfeld phase, we have
\begin{equation}\label{gamma}
\begin{aligned}
\dot{\kappa}_N&=-i\langle\epsilon_N|\partial_t|\epsilon_N\rangle+\langle\epsilon_N|H|\epsilon_N\rangle\\
&=-i\left(-iN\frac{\dot{\alpha}\cos\beta}{2}\right)+N\frac{\Delta\cos\beta}{2}\\
&+N(g_R\sin\beta\cos\alpha+g_I\sin\beta\cos\alpha) \\
&=N\frac{g_R\cos\alpha+g_I\sin\alpha}{\sin\beta}=N\dot{\kappa},
\end{aligned}
\end{equation}
with $\dot{\kappa}\equiv(g_R\cos\alpha+g_I\sin\alpha)/\sin\beta$, in which we have applied the time-derivative of the operators $A$ and $B$
\begin{equation}
\begin{aligned}
\frac{\partial}{\partial t}A^\dagger&=-\frac{\dot{\beta}}{2}B^\dagger-\frac{i\dot{\alpha}}{2}\left(\cos\beta A^\dagger+\sin\beta B^\dagger\right),\\
\frac{\partial}{\partial t}B^\dagger&=\frac{\dot{\beta}}{2}B^\dagger-\frac{i\dot{\alpha}}{2}\left(\sin\beta A^\dagger-\cos\beta B^\dagger\right),
\end{aligned}
\end{equation}
and Eqs.~(\ref{thetaalpha}) and (\ref{HAB}).

Given the time-dependent parameters $\beta(t)$, $\alpha(t)$, and $\kappa(t)$ in Eqs.~(\ref{thetaalpha}) and (\ref{gamma}), the coupling strengths and the effective frequency of the hybrid mode that are directly relevant in quantum control can be expressed by
\begin{equation}\label{control}
\begin{aligned}
g_R&=\dot{\kappa}\cos\alpha\sin\beta-\frac{\dot{\beta}}{2}\sin\alpha,\\
g_I&=\dot{\kappa}\sin\alpha\sin\beta+\frac{\dot{\beta}}{2}\cos\alpha,\\
\Delta&=\dot{\alpha}+2\dot{\kappa}\cos\beta.
\end{aligned}
\end{equation}
By virtue of their excitation-number-independence, the transfer of an arbitrary superposed state from mode-$m$ to mode-$b$ can be achieved under the boundary conditions $\beta(0)=0$ and $\beta(T)=\pi$. Thus in a general situation, an initial state
\begin{equation}
|\psi(0)\rangle=\sum_kC_k|k0\rangle=\sum_kC_k|\psi_k(0)\rangle,
\end{equation}
where $C_k$ is the time-independent normalized coefficients, will evolve to
\begin{equation}
|\psi(T)\rangle=\sum_{k}C_k e^{-ik(\kappa+\alpha)}|0k\rangle,
\end{equation}
at the final time $T$, according to Eqs.~(\ref{phit}) and (\ref{operatorABLr}).

\begin{figure}[htbp]
\centering
\includegraphics[width=0.23\textwidth]{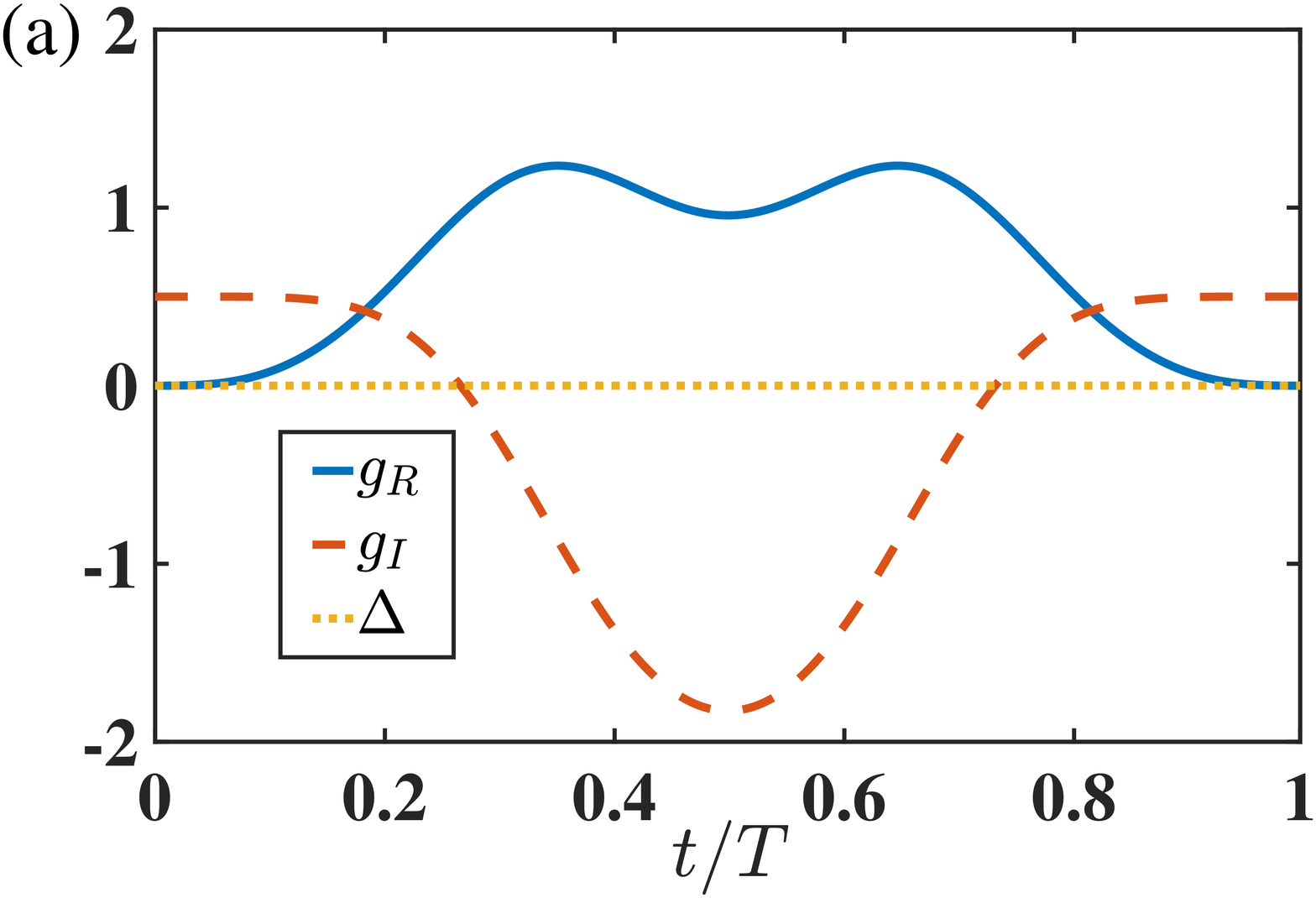}
\includegraphics[width=0.23\textwidth]{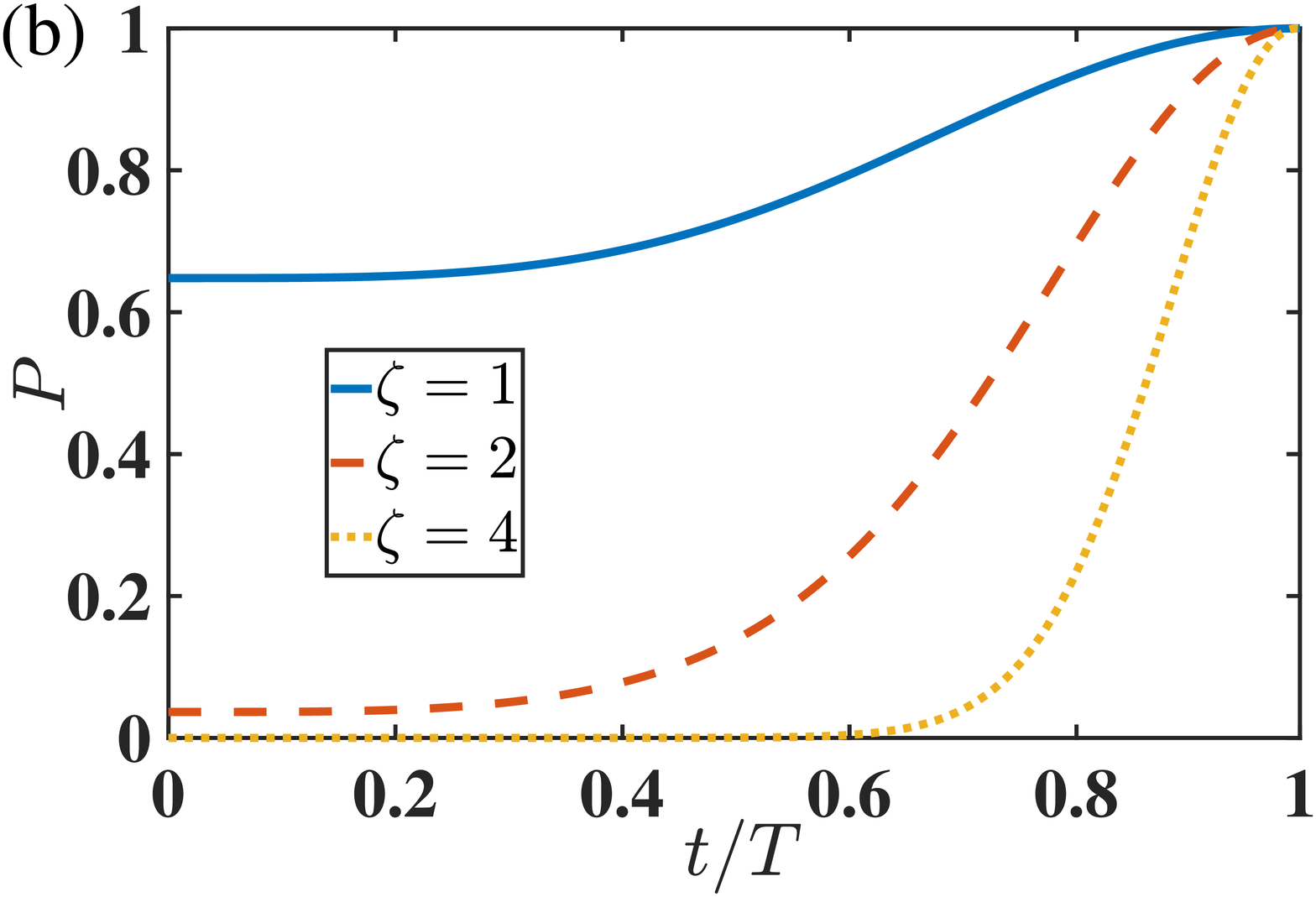}
\caption{(a) The shapes of the real and imaginary parts of the driving-enhanced coupling strength $g_R$ and $g_I$ and the effective frequency of the hybrid mode $\Delta$, in units of the coupling strength $\Omega$ used in Eq.~(\ref{gGDelta1}). (b) The dynamics of the state population of mode-$b$ under various initial cat states of the hybrid mode-$m$. Here the control parameters are set as $\beta=\pi t/T$, $\alpha=-4/3\sin^3\beta$, and $\kappa=\beta-\sin(2\beta)/2$.}\label{catlr}
\end{figure}

The state transfer assisted by the Lewis-Riesenfeld invariant can be verified in Fig.~\ref{catlr} by the state population $P$ of the phonon mode-$b$ given by Eq.~(\ref{population}). With the selected control functions of $\beta$, $\alpha$, and $\kappa$, one can directly find the time-dependence of the real and imaginary parts of the coupling strength $g_R$ and $g_I$ and the frequency of the hybrid mode $\Delta$ through Eq.~(\ref{control}). They are plotted in Fig.~\ref{catlr}(a). The initial states in Fig.~\ref{catlr}(b) are various cat states with $\zeta=1,2,4$. It is found that a perfect transfer can always be achieved via the LR-invariant-based inverse engineering.

\section{State transfer under systematic errors}\label{syserror}

In practice the ideal trajectory of the control parameters is not implemented exactly because of technical imperfections and constraints. These systematic errors pose the need for studying the effect of perturbations on transport protocols and optimizing protocols that are robust with respect to the stochastic fluctuation in the Hamiltonian~\cite{syserror}. In this section, the Hamiltonian implemented in experiments could be assumed to be
\begin{equation}\label{Htotalerror}
H_{\rm exp}=H(t)+\gamma H_g+\eta H_{\Delta},
\end{equation}
where $H(t)=H_g+H_{\Delta}$ is the ideal or unperturbed Hamiltonian in Eq.~(\ref{Hamiltonian}), $H_g$ and $H_{\Delta}$ are respectively the interaction Hamiltonian between the hybrid mode and the phonon mode and the bare Hamiltonian for them, i.e.,
\begin{equation}\label{Herr}
\begin{aligned}
H_{g}&\equiv g(t)m^\dagger b+g^*(t)mb^\dagger,\\
H_{\Delta}&\equiv\frac{\Delta(t)}{2}(m^\dagger m-b^\dagger b).
\end{aligned}
\end{equation}
And $\gamma$ and $\eta$ are the dimensionless perturbed coefficient for the coupling strength and the frequency detuning, respectively.

With the practical Hamiltonian~(\ref{Htotalerror}), the evolved state is obtained from the Schr\"odinger equation
\begin{equation}\label{schroerr}
i\frac{\partial}{\partial t}|\Psi(t)\rangle=[H(t)+\gamma H_{g}+\eta H_{\Delta}]|\Psi(t)\rangle.
\end{equation}
Then the sensitivity to the systematic error is defined as
\begin{equation}\label{qs}
\begin{aligned}
q_g&=-\frac{\partial{P(T)}}{\partial(\gamma^2)}\bigg|_{\gamma=0},\\
q_{\Delta}&=-\frac{\partial{P(T)}}{\partial(\eta^2)}\bigg|_{\eta=0}.
\end{aligned}
\end{equation}
where $P(T)$ is the state population evaluated by replacing the evolved state in Eq.~(\ref{population}) with $|\Psi(T)\rangle$ at the final time $T$. We write $P(T)$ as $P$ for simplicity in the subsequent subsections~\ref{pipulse}, \ref{Tqderror}, and \ref{lrerror}.

\subsection{$\pi$-pulse}\label{pipulse}

Before working on the two accelerated adiabatic passages, we first consider a straightforward $\pi$-pulse for the state transfer. Then in Eq.~(\ref{Hamiltonian}), it is found that $\Delta(t)=0$ and $\int^T_0dtg(t)=\pi/2$. Accordingly, Eq.~(\ref{Htotalerror}) becomes $H_{\rm exp}=(1+\gamma)H_g$, which can be diagonalized with the new operators $A=(m+b)/\sqrt{2}$ and $B=(m-b)/\sqrt{2}$, similarly to the transformation in Eqs.~(\ref{ham}) and (\ref{AB}). The special initial state $|\Psi(0)\rangle=|N\rangle_m|0\rangle_b=|N0\rangle$ could then expand by the eigenstates in Eq.~(\ref{En})
\begin{equation}
|N0\rangle=\frac{1}{\sqrt{2^N}}\sum^{N}_{n=0}\sqrt{C^n_N}|\epsilon_{N-n}\rangle.
\end{equation}
The eigenvalue of $|\epsilon_{N-n}\rangle$ is now $(N-2n)(1+\gamma)g(t)$ with respect to $H_{\rm exp}$. So that we have
\begin{equation}
\begin{aligned}
|\Psi(T)\rangle=\frac{1}{\sqrt{2^N}}\sum^N_{n=0}\sqrt{C^n_N}e^{-i(N-2n)(1+\gamma)\frac{\pi}{2}}|\epsilon_{N-n}\rangle.
\end{aligned}
\end{equation}
by the Schr\"odinger equation in Eq.~(\ref{schroerr}). According to Eqs.~(\ref{ham}) and (\ref{AB}), the target state is
\begin{equation}
|0N\rangle=\frac{1}{\sqrt{2^N}}\sum^{N}_{n=0}(-1)^n\sqrt{C^n_N}|\epsilon_{N-n}\rangle.
\end{equation}
Then the state-transfer fidelity measured by the target-state population $P$ in mode-$b$ is
\begin{equation}
\begin{aligned}
P=|\langle 0N|\Psi(T)\rangle|^2=\cos^{2N}\left(\frac{\pi}{2}\gamma\right).
\end{aligned}
\end{equation}
by virtue of Eq.~(\ref{population}). And by Eq.~(\ref{qs}), the systematic-error sensitivity to coupling strength for the initial Fock state $|N0\rangle$ is
\begin{equation}
q_g=\frac{N\pi^2}{4}.
\end{equation}
The preceding derivation can be straightforwardly extended to arbitrary superposed state $|\psi(0)\rangle=\sum_k C_k|k0\rangle$ by virtue of its independence of the excitation number. In general situations, we have
\begin{equation}
\begin{aligned}
&P=\sum_{C_k\neq0}|C_k|^2\cos^{2k}\left(\frac{\pi}{2}\gamma\right), \\
&q_g=\frac{\pi^2}{4}\sum_{C_k\neq0}k|C_k|^2=\frac{\pi^2}{4}\bar{n}_m.
\end{aligned}
\end{equation}
It is interesting to find that the systematic-error sensitivity in the $\pi$-pulse protocol is proportional to the average excitation number $\bar{n}_m$ of the initial state.

\subsection{Transitionless quantum driving}\label{Tqderror}

\begin{figure}[htbp]
\centering
\includegraphics[width=0.23\textwidth]{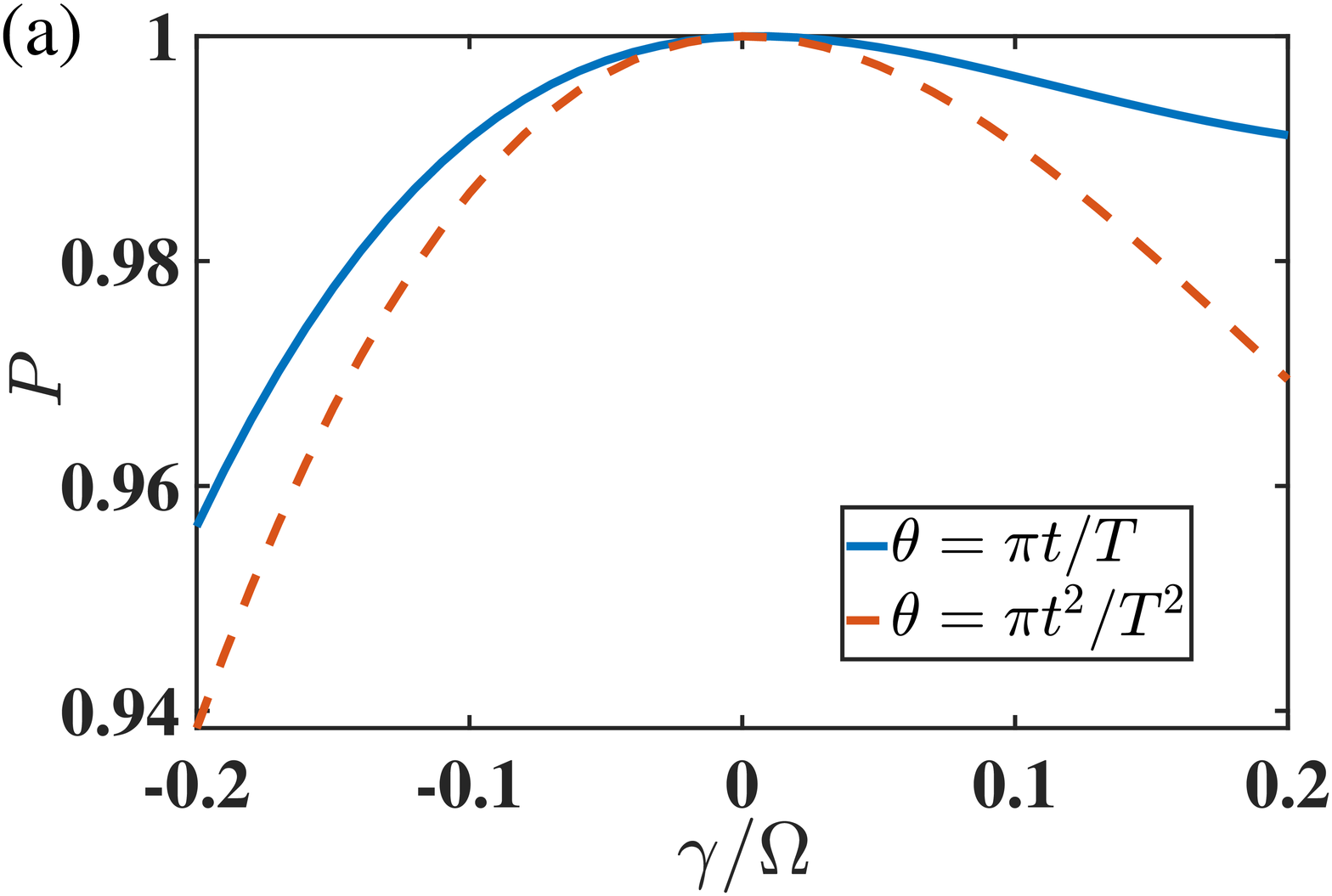}
\includegraphics[width=0.23\textwidth]{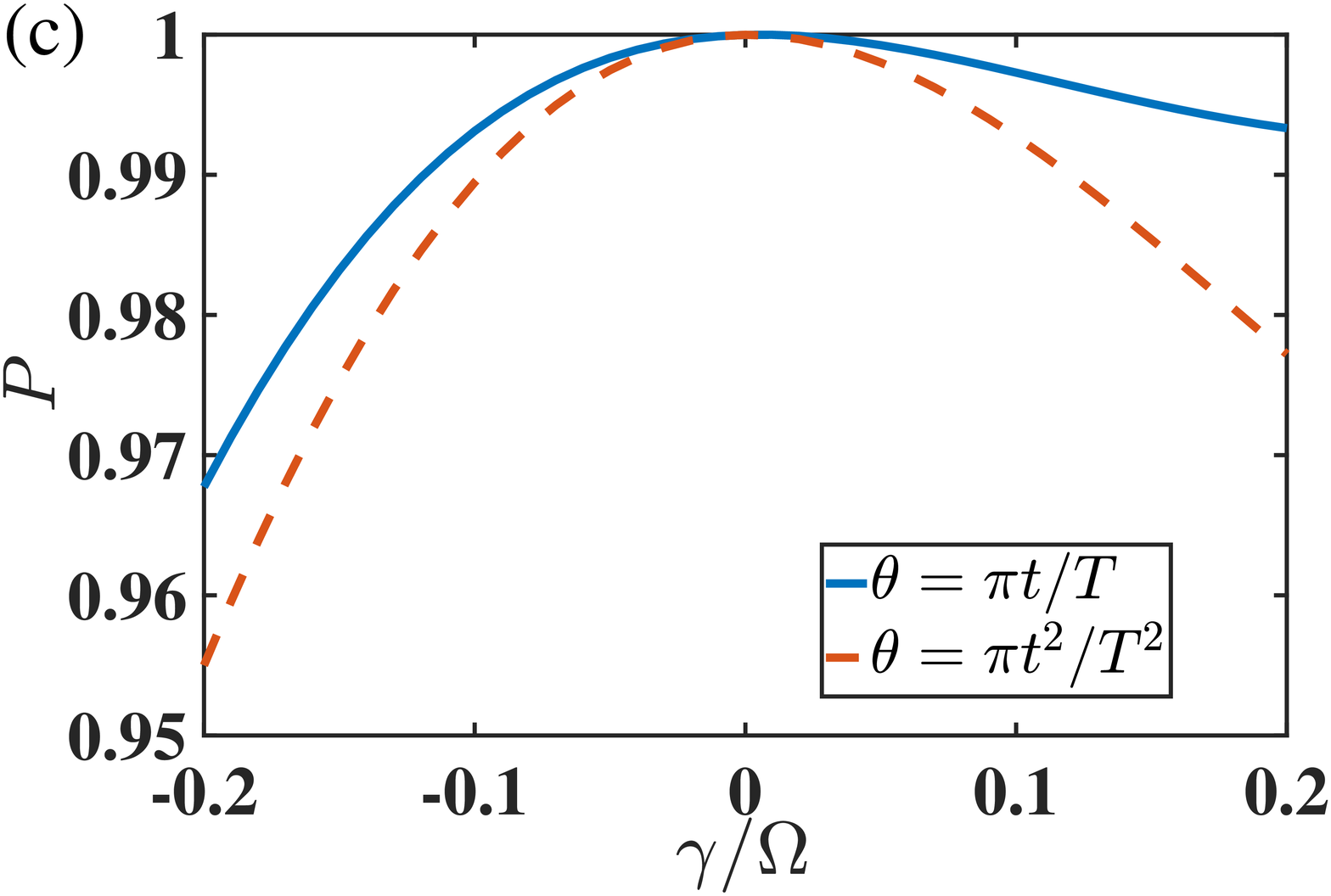}
\includegraphics[width=0.23\textwidth]{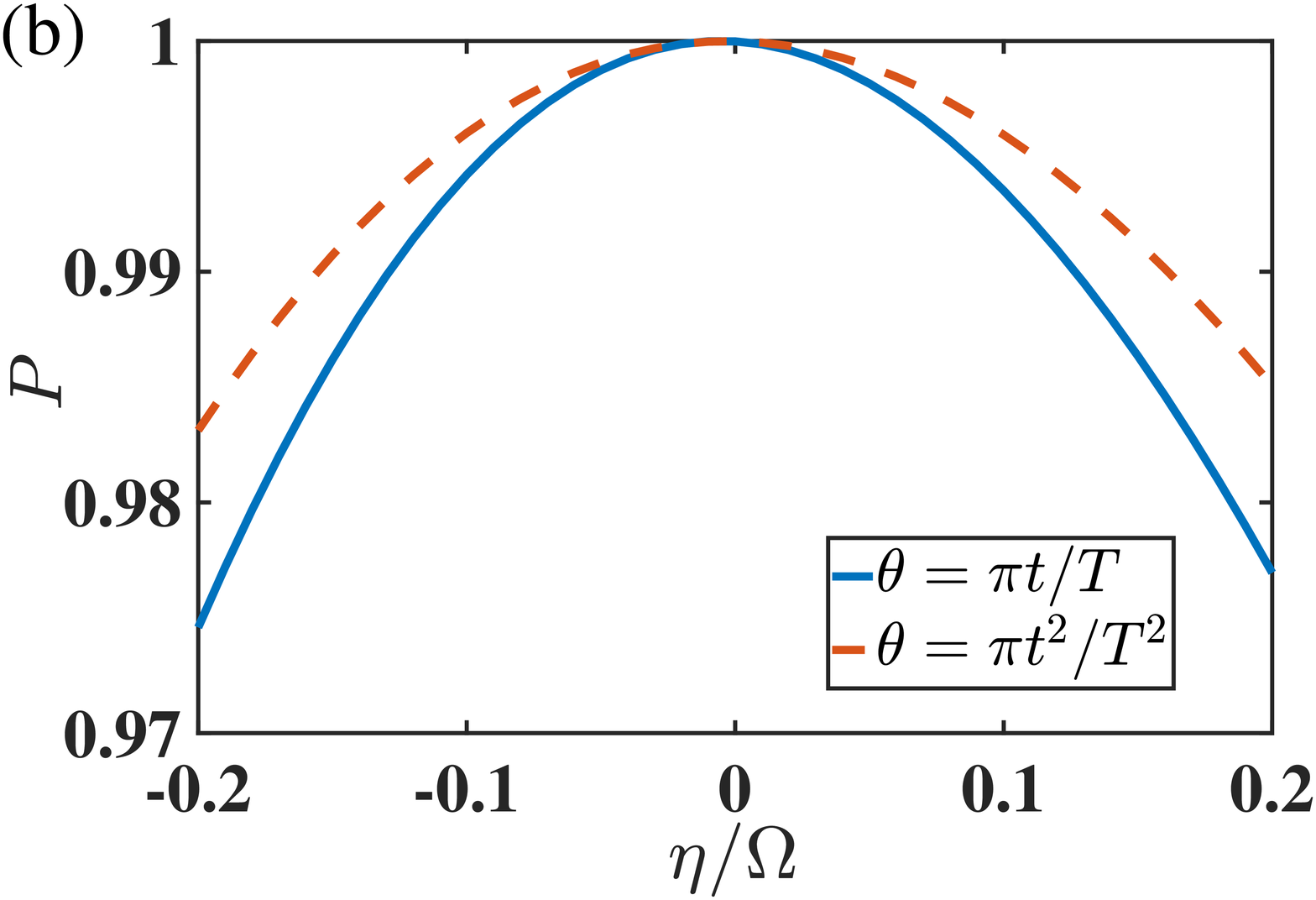}
\includegraphics[width=0.23\textwidth]{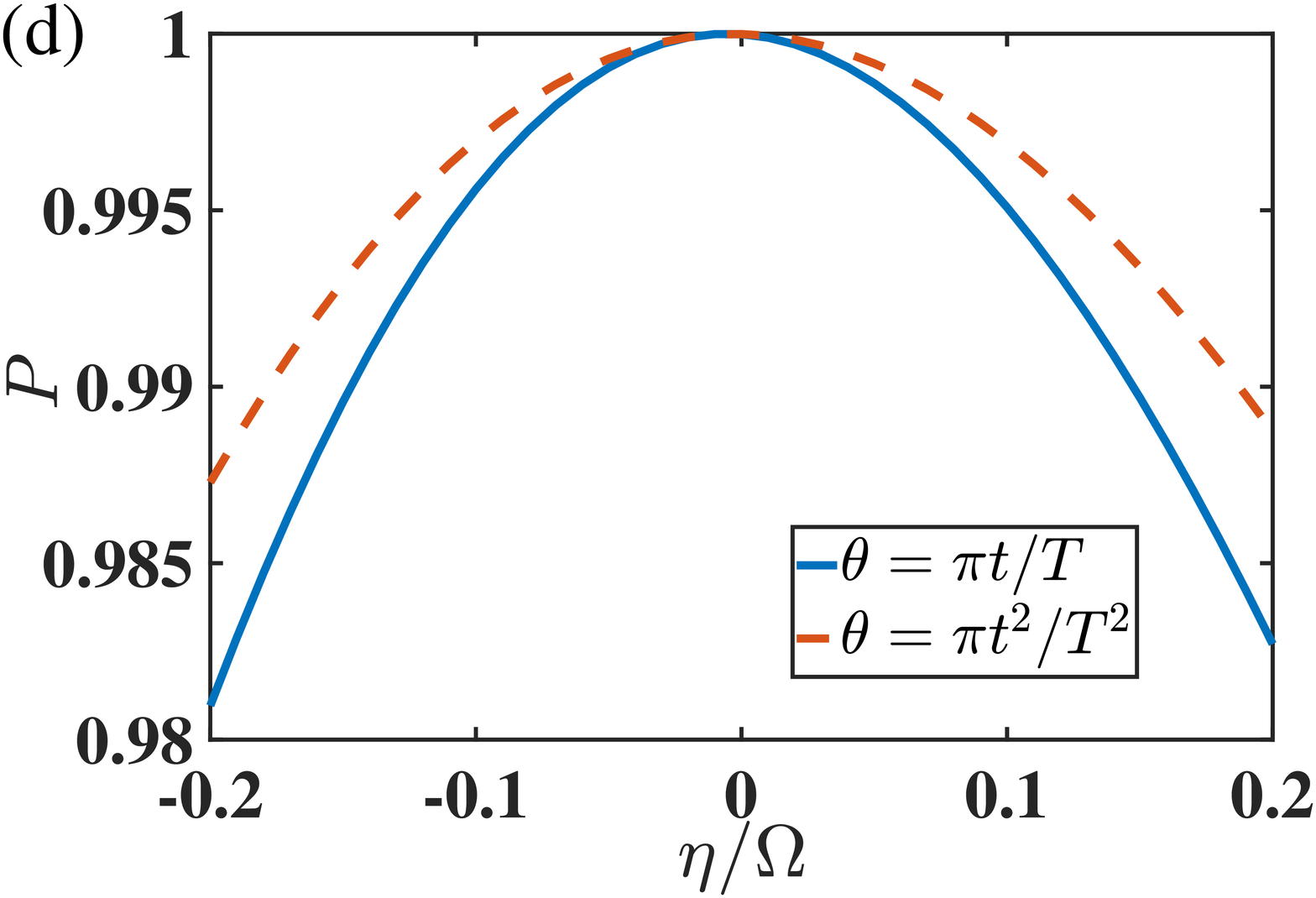}
\caption{State-transfer population $P(T)$ of the phonon mode-$b$ as a function of the systematic errors associated to coupling-strength $\gamma$ and frequency-detuning $\eta$, under various target states and shape functions of TQD protocol. In (a) and (b), the initial state is the Fock state $|1\rangle$ and in (c) and (d), it is the cat state with $\zeta=1$. $T=\pi/\Omega$. }\label{CDnoise}
\end{figure}

Now we analysis the stability for state transfer and error sensitivity of the TQD protocol provided in Sec.~\ref{secTQD}. Note the unperturbed Hamiltonian $H(t)$ in the total Hamiltonian in Eq.~(\ref{Htotalcd}) is replaced with $H_{\rm exp}$ in Eq.~(\ref{Htotalerror}). In Figs.~\ref{CDnoise}(a) and (c) $\eta=0$, and in Figs.~\ref{CDnoise}(b) and (d) $\gamma=0$. In Fig.~\ref{CDnoise}, the TQD protocols are performed by the time-dependence of the effective frequency $\Delta$ and the driving-enhanced coupling strength $g(t)$ (assumed to be real for TQD), that are determined by the shape functions of $\theta=\pi/2(t/T)$ in Eq.~(\ref{gGDelta1}) or $\theta=\pi/2(t/T)^2$. One can observe that the protocol stability is not sensitive to the choice of the target states. It is found that the impact of the coupling-strength fluctuation of the interaction Hamiltonian $H_g$ is asymmetrical to the parameter $\gamma$ in the negative and positive axis. With the same magnitude, the decrement in the state population $P$ induced by a positive $\gamma$ is clearly smaller than that by a negative $\gamma$. In particular, $P=0.99$ for $\gamma/\Omega=0.2$ and $P=0.96$ for $\gamma/\Omega=-0.2$. In contrast, the state-transfer population is roughly symmetrical to the energy fluctuation $\eta$ of the free Hamiltonian $H_{\Delta}$. Another difference between Figs.~\ref{CDnoise}(a), (c) and Figs.~\ref{CDnoise}(b), (d) manifests in the error sensitivity to the shape of the control parameter $\theta(t)$. For a nonvanishing $\gamma$ ($\eta$), the protocol is more robust with the linear function $\theta=\pi/2(t/T)$ [the quadratic function $\theta=\pi/2(t/T)^2$] than that with the quadratic function $\theta=\pi/2(t/T)^2$ [the linear function $\theta=\pi/2(t/T)$].

\begin{figure}[htbp]
\centering
\includegraphics[width=0.40\textwidth]{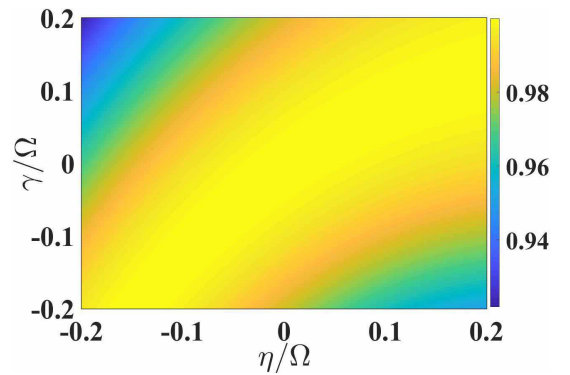}
\caption{State-transfer population $P(T=\pi/\Omega)$ of the phonon mode-$b$ in the space of the systematic-error parameters $\gamma$ and $\eta$. The target state is chosen as the cat state with $\zeta=1$. }\label{catCDnoiseboth}
\end{figure}

In Fig.~\ref{catCDnoiseboth}, we switch simultaneously on the systematic error in both interaction Hamiltonian and free Hamiltonian and fix the target state as a cat state with $\zeta=1$. Much to our anticipation, a nearly unit state-transfer is found in a remarkable regime with $\gamma\approx\eta$. It could be readily understood that when $\gamma\approx\eta$, the experimental Hamiltonian $H_{\rm exp}$ in Eq.~(\ref{Htotalerror}) is approximated by $(1+\gamma)H(t)$, equivalent to a rescaling over the whole original Hamiltonian that renders the same counter-diabatic Hamiltonian $H_{CD}$ in Eq.~(\ref{Hcdss}).

From both Figs.~\ref{CDnoise} and \ref{catCDnoiseboth}, with even $20\%$ fluctuations in system parameters, the state-transfer fidelity could be maintained above $0.95$. One can generally find that the TQD approach is robust against the systematic errors.

\subsection{Invariant-based inverse engineering}\label{lrerror}

In this subsection, we construct an optimal protocol against the systematic errors by using the Lewis-Riesenfeld invariant in Eq.~(\ref{Lr}). Now we employ the ideal Hamiltonian $H(t)$ in Eq.~(\ref{HgG}) with a complex coupling strength $g(t)$.

From the Schr\"odinger equation in Eq.~(\ref{schroerr}), one can obtain the time-evolved state at the final time $T$ up to the second-order of $O(\gamma^2)$ and $O{(\eta^2)}$:
\begin{equation}\label{PsiT2}
\begin{aligned}
&|\Psi(T)\rangle=|\psi(T)\rangle-i\gamma\int^T_0 dt U_0(T,t)H_{g}|\psi(t)\rangle\\
&-\gamma^2\int^T_0dt\int^t_0dt'U_0(T,t)H_{g}(t)U_0(t,t')H_{g}(t')|\psi(t')\rangle\\
&-i\eta\int^T_0 dt U_0(T,t)H_{\Delta}|\psi(t)\rangle\\
&-\eta^2\int^T_0dt\int^t_0dt'U_0(T,t)H_{\Delta}(t)U_0(t,t')H_{\Delta}(t')|\psi(t')\rangle\\
&+\cdots,
\end{aligned}
\end{equation}
where $|\psi(t)\rangle$ is the unperturbed solution determined by the ideal evolution operator $U_0(t,0)$. Under the assumption that the initial state is a Fock state $|N\rangle_m|0\rangle_b$ and by virtue of Eqs.~(\ref{phi}) and (\ref{phit}), the unperturbed solution reads
\begin{equation}\label{varphit}
|\psi(t)\rangle=|\epsilon_N(t)\rangle e^{-iN\kappa(t)},
\end{equation}
where $\kappa$ is defined in the last line of Eq.~(\ref{gamma}), and we have
\begin{equation}\label{U0st}
U_0(s,t)=\sum^N_{n=0}e^{-i(N-2n)[\kappa(s)-\kappa(t)]}|\epsilon_{N-n}(s)\rangle\langle\epsilon_{N-n}(t)|.
\end{equation}
Substituting Eqs.~(\ref{varphit}) and (\ref{U0st}) to Eq.~(\ref{PsiT2}), the final state population is
\begin{equation}
\begin{aligned}
P&=|\langle 0N|\Psi(T)\rangle|^2=|\langle\psi(T)|\Psi(T)\rangle|^2 \\
&\approx 1-\gamma^2\sum^N_{n=1}\left|\int^T_0dte^{-2ni\kappa(t)}\langle\epsilon_{N-n}(t)|H_g|\epsilon_N(t)\rangle\right|^2\\
&-\eta^2\sum^N_{n=1}\left|\int^T_0dte^{-2ni\kappa(t)}\langle\epsilon_{N-n}(t)|H_\Delta|\epsilon_N(t)\rangle\right|^2.
\end{aligned}
\end{equation}
By virtue of
\begin{equation}
\begin{aligned}
\langle\epsilon_{N-1}(t)|H_g|\epsilon_N(t)\rangle&=-\sqrt{N}\dot{\kappa}\sin\beta\cos\beta+\frac{i\sqrt{N}\dot{\beta}}{2},\\
\langle\epsilon_{N-n}(t)|H_g|\epsilon_N(t)\rangle&=0, \quad n\ne1\\
\langle\epsilon_{N-1}(t)|H_\Delta|\epsilon_N(t)\rangle&=\frac{\sqrt{N}\dot{\alpha}}{2}\sin\beta+\sqrt{N}\dot{\kappa}\sin\beta\cos\beta,\\
\langle\epsilon_{N-n}(t)|H_\Delta|\epsilon_N(t)\rangle&=0, \quad n\ne1
\end{aligned}
\end{equation}
and the boundary condition $\beta(0)=0$ and $\beta(0)=\pi$, the systematic error sensitivities for $|\psi(0)\rangle=|N0\rangle$ are found to be
\begin{equation}\label{sensq}
\begin{aligned}
q_g&=N\left|\int^T_0 dt\dot{\beta}\sin^2\beta e^{-2i\kappa}\right|^2,\\
q_{\Delta}&=N\left|\int^T_0dt\sin\beta\left(\frac{\dot{\alpha}}{2}+\dot{\kappa}\cos\beta\right)e^{-2i\kappa}\right|^2,
\end{aligned}
\end{equation}
according to their definitions in Eq.~(\ref{qs}). Specially when $\kappa$ is constant, we restore the result in the $\pi$-pulse case $q_g=N\pi^2/4$, which is independent of $\beta(t)$.

It is interesting to find that $q_g=0$ can be attained when
\begin{equation}\label{qg0}
\kappa(t)=j\left[\beta-\frac{\sin(2\beta)}{2}\right]
\end{equation}
with a nonzero integer $j$. And $q_{\Delta}=0$ can be attained when
\begin{equation}\label{qD0}
\frac{\dot{\alpha}}{2}+\dot{\kappa}\cos\beta=0.
\end{equation}
Equations~(\ref{qg0}) and (\ref{qD0}) render $\dot{\alpha}=-4j\dot{\beta}\cos^2\beta\sin\beta$. An immediate choice is $\alpha=-4j\cos^3\beta/3$. Therefore the optimized invariant-based inverse engineering that is robust against the systematic errors are described by
\begin{equation}\label{optimalpulse}
\begin{aligned}
&\beta(0)=0, \quad \beta(T)=\pi,\\
&\kappa(t)=j\left[\beta-\frac{\sin(2\beta)}{2}\right],\\
&\alpha(t)=-\frac{4j\cos^3\beta}{3}.
\end{aligned}
\end{equation}
Consequently, the control protocol in Eq.~(\ref{control}) turns out to be
\begin{equation}\label{optimalpulseshape}
\begin{aligned}
&g_R=2\dot{\beta}\sin^3\beta\cos\left(\frac{4}{3}\sin^3\beta\right)+\frac{\dot{\beta}}{2}\sin\left(\frac{4}{3}\sin^3\beta\right),\\
&g_I=-2\dot{\beta}\sin^3\beta\sin\left(\frac{4}{3}\sin^3\beta\right)+\frac{\dot{\beta}}{2}\cos\left(\frac{4}{3}\sin^3\beta\right),\\
&\Delta=0.
\end{aligned}
\end{equation}
where we have set $j=1$. It is more important to find that the optimized protocol in Eq.~(\ref{optimalpulseshape}) as well as the parametric setting in Eq.~(\ref{optimalpulse}) is independent of the excitation number. So that it applies to the general superposed state $|\psi(0)\rangle=\sum_kC_k|k0\rangle$, where the population is found to be
\begin{equation}
P=1-\gamma^2\sum_{C_k\neq0}|C_k|^2\left|\int^T_0dt\langle\epsilon_{k-1}(t)|H_g|\epsilon_k(t)\rangle\right|^2.
\end{equation}
Note in the optimized control by Eq.~(\ref{optimalpulseshape}), $\Delta=0$ implies that $P$ is strictly insensitive to $\eta$.

\begin{figure}[htbp]
\centering
\includegraphics[width=0.23\textwidth]{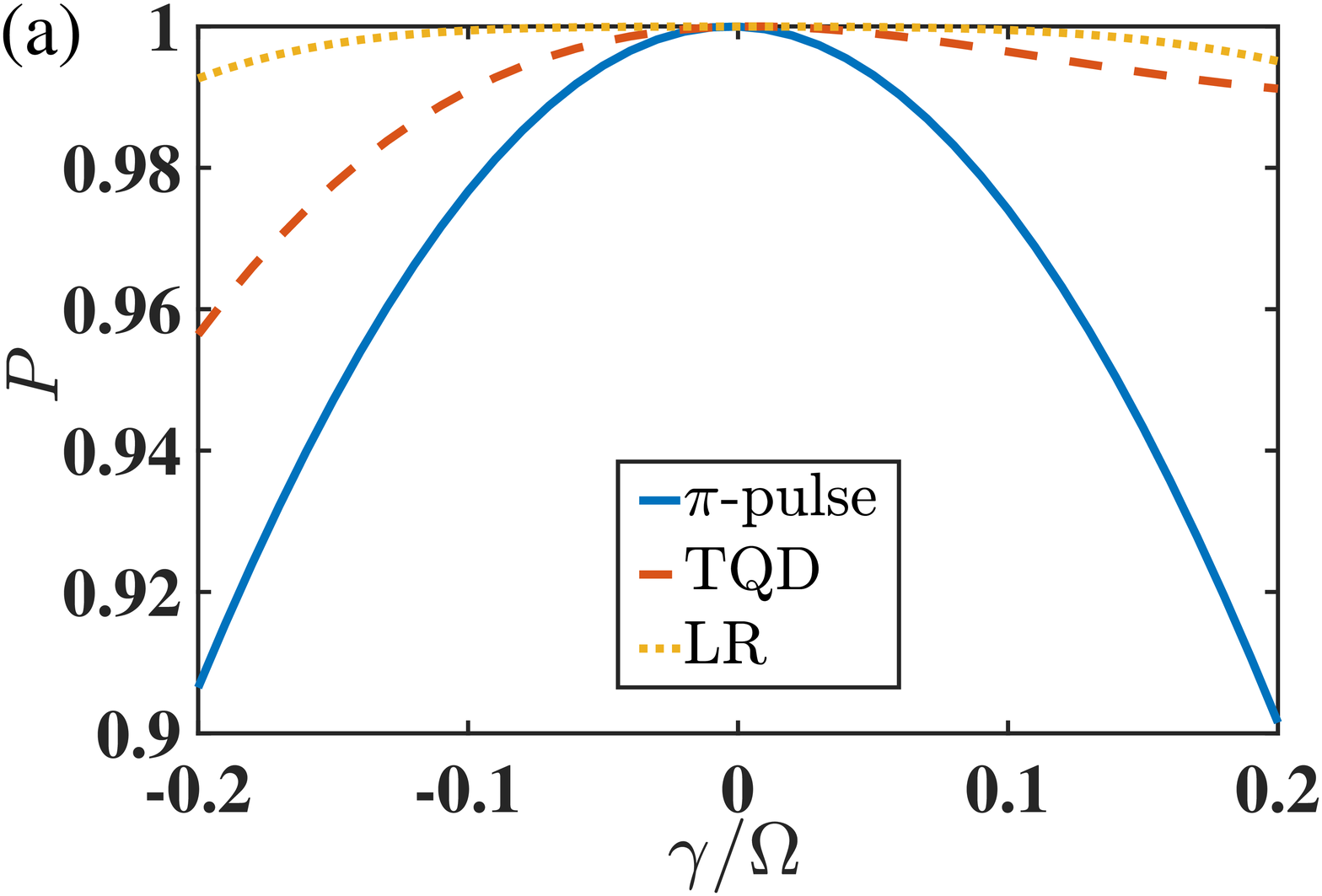}
\includegraphics[width=0.23\textwidth]{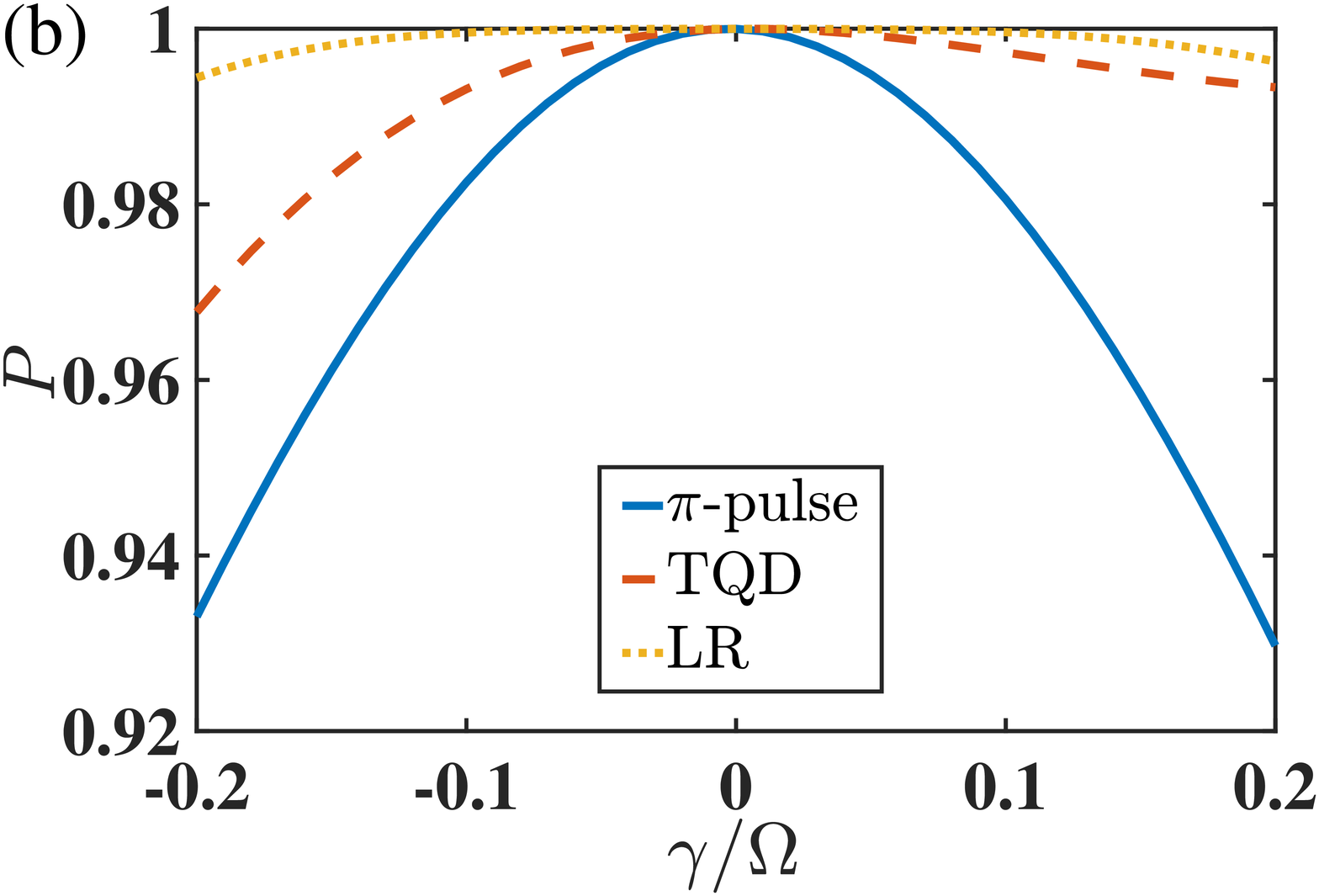}
\caption{State-transfer population $P(T=\pi/\Omega)$ of the phonon mode-$b$ as a function of the systematic errors associated to coupling-strength $\gamma$ under various protocols. (a) the initial state is the Fock state $|1\rangle$, (b) it is the cat state with $\zeta=1$.}\label{noise}
\end{figure}

In Fig.~\ref{noise}, we compare the systematic-error sensitivity in various state-transfer protocols, including the flat $\pi$ pulse (see the blue solid lines), the transitionless quantum driving protocol described by the parametric functions in Eq.~(\ref{gGDelta1}) (see the red dashed lines), and the optimized protocol based on the LR-invariant described by Eq.~(\ref{optimalpulseshape}) (see the orange dotted lines). It is found for both Fock-state and cat state, the optimized protocol assisted by the LR-invariant demonstrates a much stronger robustness than the other two protocols. The population is maintained as unit within a significant regime of fluctuations of $10\%$ magnitude in terms of the coupling strength. The flat $\pi$ pulse behaves as the most fragile protocol.

\section{Discussion}\label{discussion}

Alternatively, the robustness of the state-transfer protocols can be tested by taking the effects of dissipative thermal baths into account. We now calculate the transfer population with the standard Lindblad master equation under the Born-Markovian approximation. The dynamical equation reads,
\begin{equation}
\begin{aligned}
\frac{\partial\rho(t)}{\partial t}&=-i[H(t),\rho(t)]\\
&+\left[\kappa_m(\bar{n}_m+1)L(m)+\kappa_m\bar{n}_mL(m^\dagger)\right]\rho\\
&+\left[\kappa_b(\bar{n}_b+1)L(b)+\kappa_b\bar{n}_bL(b^\dagger)\right]\rho,
\end{aligned}
\end{equation}
where the superoperation for any Lindblad operator $o$, $o=m,b$, is defined as
\begin{equation}
L(o)\rho\equiv o\rho o^\dagger-\frac{1}{2}o^\dagger o\rho-\frac{1}{2}\rho o^\dagger o.
\end{equation}
Here $\rho$ is the density operator of the two modes and $H(t)$ is the system Hamiltonian in Eq.~(\ref{Htotalcd}) for TQD protocol or that in Eq.~(\ref{HgG}) for invariant protocol. $\kappa_m$ and $\kappa_b$ represent the decay rates of the hybrid mode-$m$ and phonon mode-$b$, respectively. In the numerical evaluation, we choose the mode frequencies as $\omega_m/2\pi=10$ GHz and $\omega_b/2\pi=10$ MHz and the damping rates as $\kappa_b=100$ Hz and $\kappa_m=10$ kHz~\cite{mppentang,magnoncavity}. The two thermal baths are assumed to be at the temperature $T_{\rm th}$, then the average excitation number for the mode-$o$ is $\bar{n}_o=[\exp(\omega_o/k_BT_{\rm th})-1]^{-1}$. The evolution time is fixed as $\Omega T=\pi$ with $\Omega/2\pi=1$ MHz.

\begin{figure}[htbp]
\centering
\includegraphics[width=0.23\textwidth]{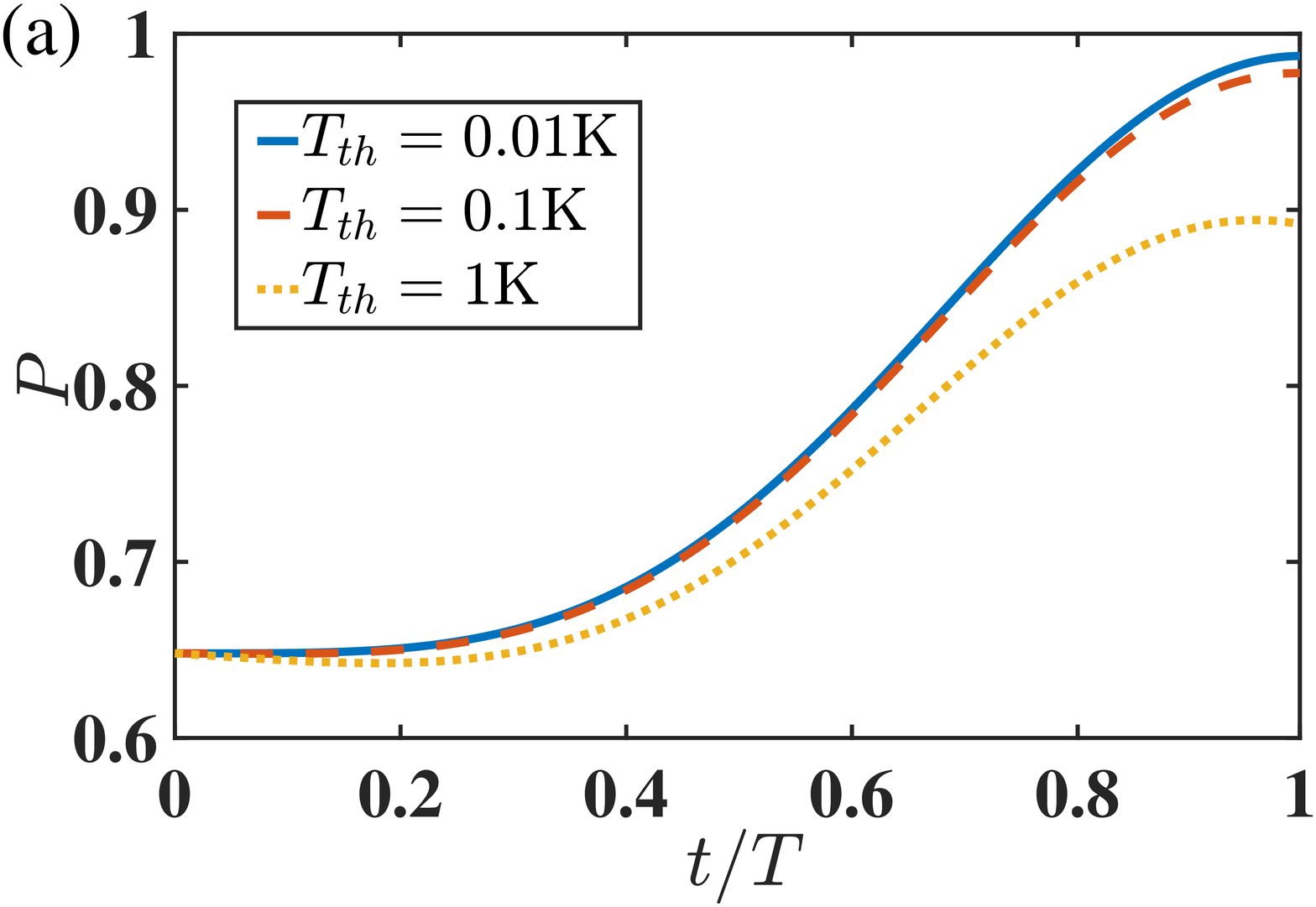}
\includegraphics[width=0.23\textwidth]{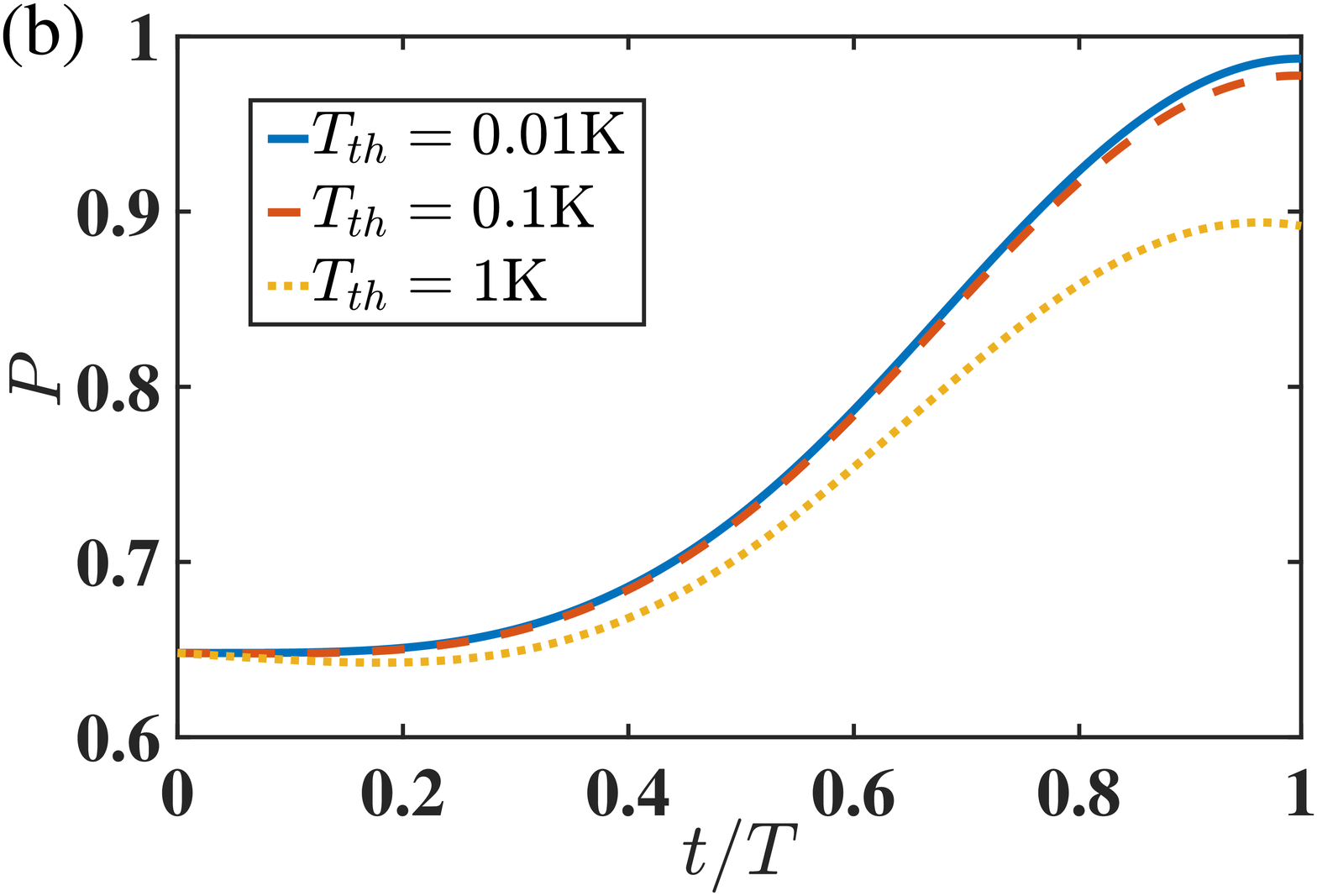}
\caption{The dynamics of target-state population in mode-$b$ in the presence of the thermal baths with various temperatures. In (a) we use the TQD protocol as described in Fig.~\ref{onecd}, and in (b) we use the LR-invariant protocol as described in Fig.~\ref{catlr}. Here the initial state is set as the cat state with $\zeta=1$.}\label{lindblad}
\end{figure}

The transfer population defined in Eq.~(\ref{population}) for a special cat state under various temperatures is plotted in Fig.~\ref{lindblad}. Comparing Figs.~\ref{lindblad}(a) and \ref{lindblad}(b), it is found that for both protocols of accelerated adiabatic passage, i.e., TQD- and invariant-STA, the effects of external thermal baths are almost the same. They are found to be robust against the thermal baths surrounding the system under low temperatures. $P$ could be maintained over $0.97$ when $T_{\rm th}\le 0.1$ K. Even under a high-temperature, e.g., $T_{\rm th}=1$ K, the population is still above $0.88$.

Our proposals are based on the system Hamiltonian~(\ref{Hamiltonian}) under the rotating-wave approximation. The ignorance of the counter-rotating terms means the coupling strength $g$ should be kept much smaller than the eigenfrequency $\omega_b$. However, $g\ll\omega_b$ might be invalid in any protocols employing accelerated adiabatic passages. To consider the effect from the counter-rotating terms, we should move back to the linearized Hamiltonian in Eq.~(\ref{Hlinearnormal}) having two hybrid modes and the phonon mode. After the unitary transformation with respect to $U=\exp[i\int_0^tds\Delta'(s)a^\dagger a]$, we have
\begin{equation}\label{Hlinearrotat}
\begin{aligned}
H&=\Delta m^\dagger m+\omega_b b^\dagger b+g(m^\dagger b+m^\dagger b^\dagger)+g^*(mb^\dagger+mb)\\
&+g'\left\{a^\dagger be^{i[\int_0^tds\Delta'(s)]}+a^\dagger b^\dagger e^{i[\int_0^tds\Delta'(s)]}\right\}\\
&+g'^*\left\{ab^\dagger e^{-i[\int_0^tds\Delta'(s)]}+abe^{-i[\int_0^tds\Delta'(s)]}\right\}.
\end{aligned}
\end{equation}
Discarding the fast-oscillating terms and using the same conventions in Eq.~(\ref{Hamiltonian0}), the Hamiltonian turns out to be
\begin{equation}\label{Hant}
\begin{aligned}
H&=[\omega_b+\Delta(t)]m^\dagger m+\omega_b b^\dagger b\\
&+g(t)m^\dagger (b+b^\dagger)+g^*(t)m(b+b^\dagger).
\end{aligned}
\end{equation}
Note it can be reduced to the Hamiltonian~(\ref{Hamiltonian}) under rotating-wave approximation with a sufficiently large $\omega_b$.

\begin{figure}[htbp]
\centering
\includegraphics[width=0.23\textwidth]{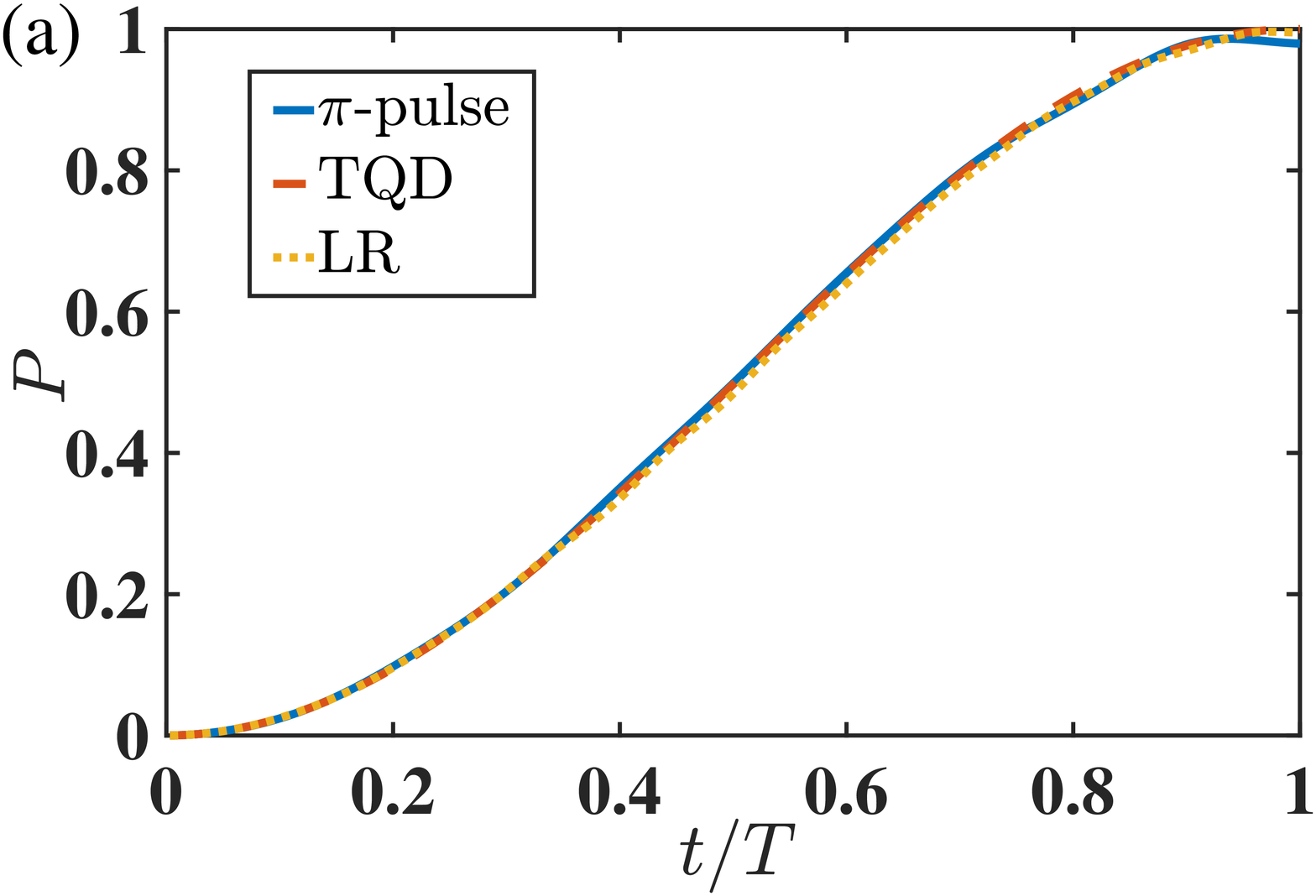}
\includegraphics[width=0.23\textwidth]{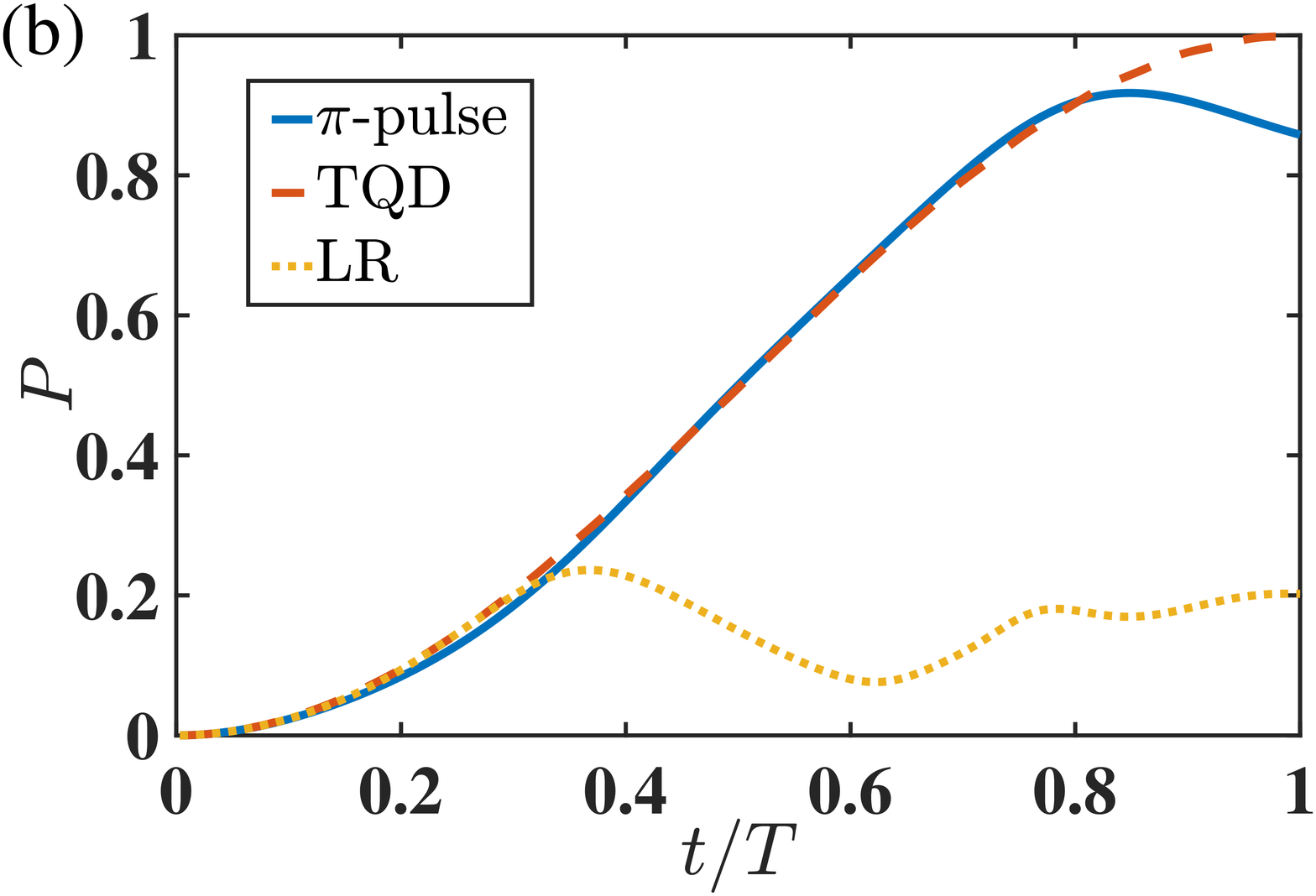}
\includegraphics[width=0.23\textwidth]{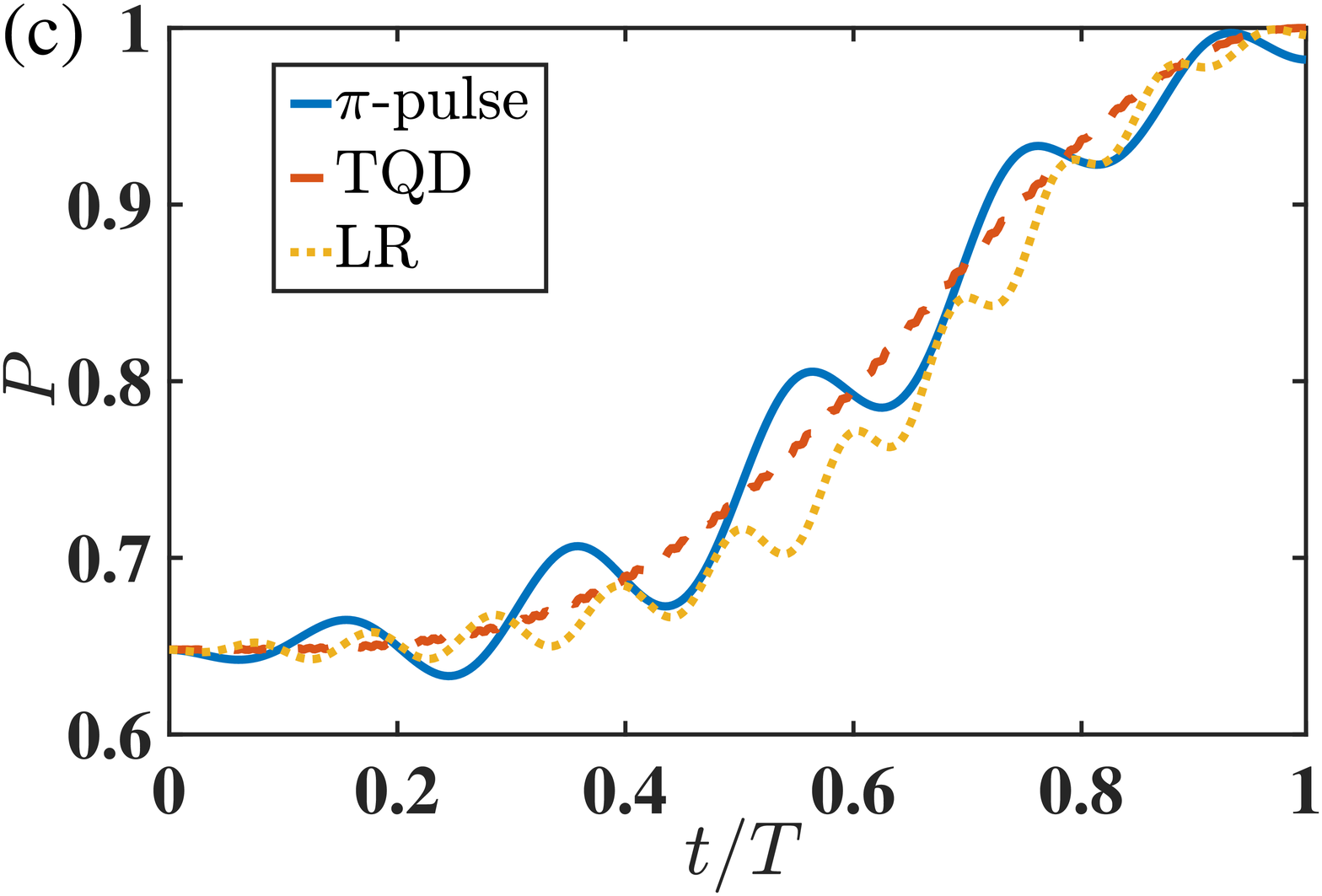}
\includegraphics[width=0.23\textwidth]{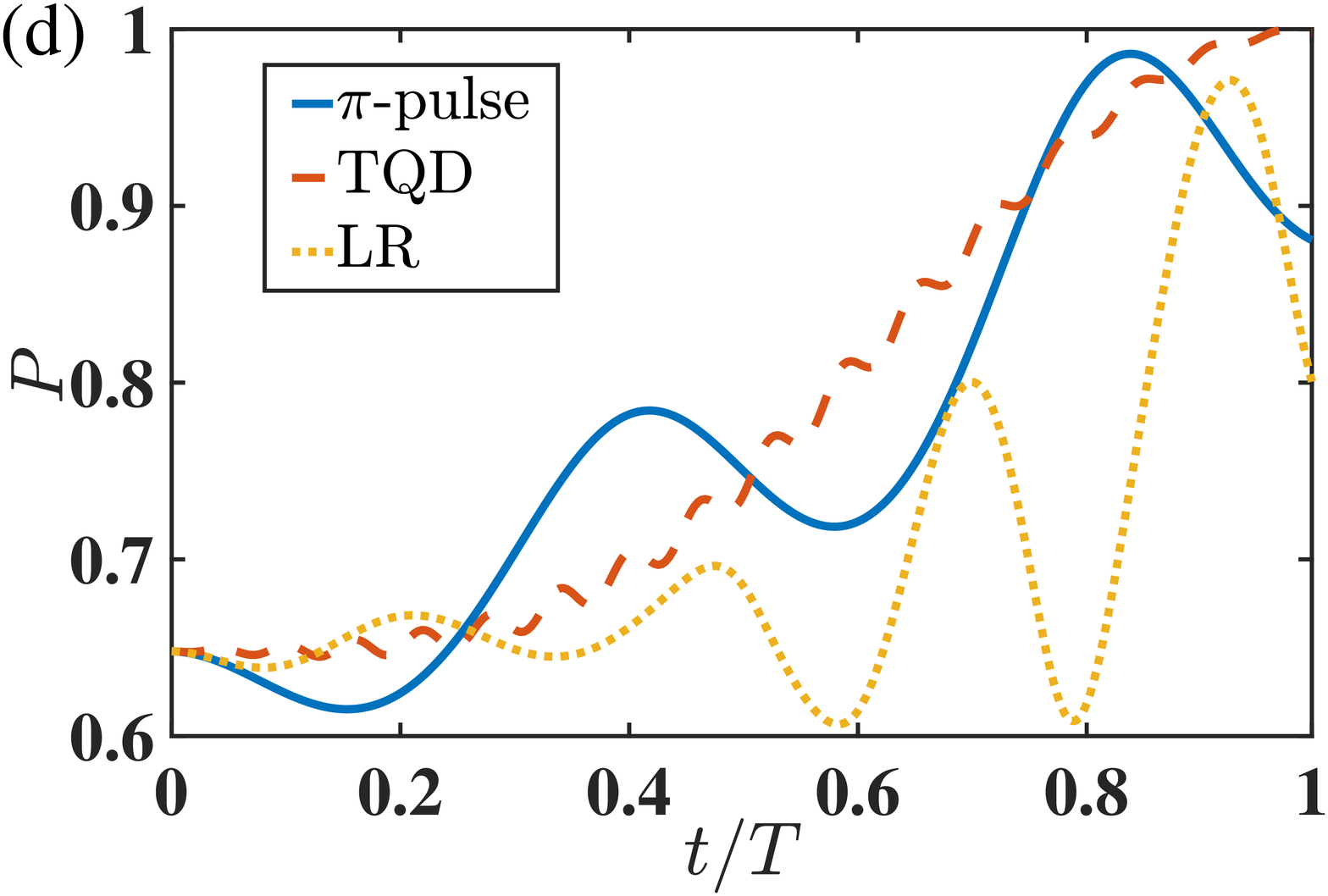}
\caption{The population dynamics in the presence of the counter-rotating interaction under various state-transfer protocols. In (a) and (b), the initial state is the Fock state $|1\rangle$ and in (c) and (d), it is the cat state with $\zeta=1$. The phonon-mode frequency is set as $\omega_b/\Omega=10$ in (a) and (c) and set as $\omega_b/\Omega=4$ in (b) and (d). $\Omega T=\pi$. }\label{antiRWA}
\end{figure}

Figure~\ref{antiRWA} demonstrates the effect of the counter-rotating terms under various transfer protocols and target states. The TQD protocol and the invariant protocol use the same groups of parameters as in Fig.~\ref{onecd}(a) and Fig.~\ref{catlr}(a), respectively. And the transfer populations for the Fock state $|10\rangle$ and the superposed state, i.e., the cat state of $|\zeta=1\rangle$, are presented respectively in Figs.~\ref{antiRWA}(a), (b) and Figs.~\ref{antiRWA}(c), (d). It is found in Figs.~\ref{antiRWA}(a) and (c) that the counter-rotating Hamiltonian could be omitted with a sufficiently large $\omega_b$. When $\omega_b/\Omega=10$, the population under the $\pi$-pulse protocol would slightly deviate from the unit transfer. It is interesting to find in Figs.~\ref{antiRWA}(b) and (d) that with a smaller $\omega_b/\Omega=4$, the invariant-based protocol [under the optimized condition indicated by Eq.~(\ref{optimalpulseshape})] yields the most disappointing behavior among the three protocols. Especially when the initial state is the Fock state $|10\rangle$, $P$ becomes even smaller than $0.2$. And the performance of the TQD protocol is still perfect, showing a great robustness to the presence of the counter-rotating interactions.

\section{Conclusion}\label{conclusion}

In this work, we have presented two popular shortcut-to-adiabatiticy protocols to realize a general state transfer in the cavity magnomechanical system, where the magnon mode simultaneously coupled to a microwave cavity mode and the mechanical-vibration mode in the same YIG sphere. Particularly, our work demonstrates how to construct the counterdiabatic Hamiltonian in terms of the creation and annihilation operators instead of the system-eigenstates and their time-derivatives, which generally applies to the continuous-variable systems. Also we derives in details the Levis-Riesenfeld invariant, based on which an inverse engineering for arbitrary initial state could be carried on. Through the analysis over the systematic error in various protocols, we obtain an optimized regime for TQD protocol and an optimized parametric setting for LR-invariant-based protocol. With the same systematic error in coupling strength, the LR protocol is superior to the TQD protocol. While TQD protocol is found to accommodate the counter-rotating interactions. Both of them are robust against the external thermal noises.

Our work in pursuit of the quantum state transfer and protection provides an important application of the cavity-magnomechanical system as a promising hybrid platform for quantum information processing. Also it applies to constructing the accelerated adiabatic passages in a general continuous-variable system.

\section*{Acknowledgments}

We acknowledge grant support from the National Science Foundation of China (Grants No. 11974311 and No. U1801661).

\appendix

\section{Effective Hamiltonian for the hybrid magnomechanical system}\label{appa}

This appendix contributes to deriving the effective Hamiltonian in Eq.~(\ref{Hlinear}). With respect to the transformation $U=\exp[i\omega_pt(a_1^\dagger a_1+m_1^\dagger m_1)]$, the original Hamiltonian in Eq.~(\ref{Hmodel}) turns out to be
\begin{equation}\label{Hmodels}
\begin{aligned}
H_0&=\Delta_a a_1^\dagger a_1+\Delta_m m_1^\dagger m_1+\omega_b b_1^\dagger b_1+g_{ma}(a_1m_1^\dagger+a_1^\dagger m_1)\\
&+g_{mb}m_1^\dagger m_1(b+b^\dagger)+i(\epsilon_p a_1^\dagger-\epsilon^*_p a_1),
\end{aligned}
\end{equation}
where $\Delta_a=\omega_a-\omega_p$ and $\Delta_m=\omega_m-\omega_p$. In the strong coupling regime for the magnon-photon interaction, we have $g_{ma}\gg\kappa_1, \kappa_2$, where $\kappa_1$ and $\kappa_2$ represent the decay rates of the photon and magnon, respectively.
The magnon mode, i.e., the collective spin-wave excitations, can efficiently interface with microwave photons, thereby consolidating the strength of dispersive interaction to produce well-separated hybridized states. These dressed normal modes read,
\begin{equation}\label{normalmode}
\begin{aligned}
m=\sin\phi a_1-\cos\phi m_1,\\
a=\cos\phi a_1+\sin\phi m_1, \\
\end{aligned}
\end{equation}
where $\tan(2\phi)\equiv2g_{ma}/{(\Delta_a-\Delta_m)}$ and $\phi\in[0,\pi/2]$. Then the Hamiltonian of the hybrid photon-magnon-phonon system in Eq.~(\ref{Hmodels}) can be rewritten as
\begin{equation}\label{Hmodelnormal}
\begin{aligned}
H&=\Delta' a^\dagger a+\Delta m^\dagger m+\omega_b b^\dagger b\\
&+i\epsilon_p(\cos\phi a^\dagger+\sin\phi m^\dagger)-i\epsilon_p^*(\cos\phi a+\sin\phi m)\\
&+g_{mb}(b+b^\dagger)(\sin^2\phi a^\dagger a-\sin\phi\cos\phi a^\dagger m\\
&-\sin\phi\cos\phi a m^\dagger +\cos^2\phi m^\dagger m),
\end{aligned}
\end{equation}
with
\begin{equation}\label{normalvalue}
\begin{aligned}
\Delta=\frac{\omega_a+\omega_m}{2}-\omega_p-\sqrt{\left(\frac{\omega_a-\omega_m}{2}\right)^2+g^2_{ma}},\\
\Delta'=\frac{\omega_a+\omega_m}{2}-\omega_p+\sqrt{\left(\frac{\omega_a-\omega_m}{2}\right)^2+g^2_{ma}}.\\
\end{aligned}
\end{equation}
Due to the input-output theory or the Heisenberg-Langevin equation, the time evolution of the system operators satisfy
\begin{equation}\label{langevinnormal}
\begin{aligned}
\dot{a}&=-(i\Delta'+\kappa_a)a+\epsilon_p\cos\phi-ig_{mb}\sin^2\phi a(b+b^\dagger)\\
&-ig_{mb}\sin\phi\cos\phi m(b+b^\dagger)+\sqrt{2\kappa_a}a_{\rm in},\\
\dot{m}&=-(i\Delta+\kappa_m)m+\epsilon_p\sin\phi-ig_{mb}\cos^2\phi m(b+b^\dagger)\\
&-ig_{mb}\sin\phi\cos\phi a(b+b^\dagger)+\sqrt{2\kappa_m}m_{\rm in},\\
\dot{b}&=-(i\omega_b+\kappa_b)b-ig_{mb}(\sin^2\phi a^\dagger a-\sin\phi\cos\phi a^\dagger m\\
&-\sin\phi\cos\phi am^\dagger +\cos^2\phi m^\dagger m)+\sqrt{2\kappa_b}b_{\rm in},
\end{aligned}
\end{equation}
where $\kappa_a=\kappa_1\cos^2\phi+\kappa_2\sin^2\phi$, $\kappa_m=\kappa_1\sin^2\phi+\kappa_2\cos^2\phi$~\cite{magnoncavity}, and $\kappa_b$ are the decay rates of the modes $a$, $m$, and $b$, respectively.

The steady-state values $a_{s}\equiv\langle a\rangle$, $m_{s}\equiv\langle m\rangle$, and $b_s\equiv\langle b\rangle$ are determined by letting $\dot{a}=\dot{m}=\dot{b}=0$. We have
\begin{equation}\label{steadynormal}
\begin{aligned}
&-(i\Delta'+\kappa_a)a_{s}+\epsilon_p\cos\phi-ig_{mb}a_{s}(b_s+b^*_s)\sin^2\phi\\
&-ig_{mb}m_{s}(b_s+b^*_s)\sin\phi\cos\phi=0,\\
&-(i\Delta+\kappa_m)m_{s}+\epsilon_p\sin\phi-ig_{mb}m_{s}(b_s+b^*_s)\cos^2\phi\\
&-ig_{mb}a_{s}(b_s+b^*_s)\sin\phi\cos\phi=0,\\
&-(i\omega_b+\kappa_b)b_s-ig_{mb}(|a_{s}|^2\sin^2\phi-a^*_{s}m_{s}\sin\phi\cos\phi\\
&-a_{s}m^*_{s}\sin\phi\cos\phi+|m_{s}|^2\cos^2\phi)=0.
\end{aligned}
\end{equation}
Due to the fact that $g_{mb}\ll\omega_b$~\cite{magnoncavity}, the last equation yields $b_s\approx 0$. And then we have
\begin{equation}\label{steadynormals}
m_{s}=\frac{\epsilon_p \sin\phi}{i\Delta+\kappa_m}, \quad
a_{s}=\frac{\epsilon_p \cos\phi}{i\Delta'+\kappa_a}.
\end{equation}

Following the standard linearization approach~\cite{optcavity}, we can rewrite the linearized Hamiltonian using the hybrid modes as
\begin{equation}\label{Hlinearnormal}
\begin{aligned}
H&=\Delta' a^\dagger a+\Delta m^\dagger m+\omega_b b^\dagger b+(g' a^\dagger+g'^*a)(b+b^\dagger)\\
&+(g m^\dagger+g^*m)(b+b^\dagger),
\end{aligned}
\end{equation}
by substituting the steady-state values in Eq.~(\ref{steadynormals}) to the Hamiltonian (\ref{Hmodelnormal}) and ignoring all the high-order terms of fluctuations and operators, and the coupling strengths are
\begin{equation}\label{Geffnormal}
\begin{aligned}
&g=g_{mb}m_{s}\cos^2\phi-g_{mb}a_{s}\sin\phi\cos\phi,\\
&g'=g_{mb}a_{s}\sin^2\phi-g_{mb}m_{s}\sin\phi\cos\phi,
\end{aligned}
\end{equation}
respectively.

According to Eq.~(\ref{normalvalue}), the two hybrid modes $a$ (with a higher frequency) and $m$ (with a lower frequency) are well well-separated polaritonic states, or normal modes, characterizing level repulsion, by the strong magnon-photon coupling $\Delta'-\Delta\ge 2g_{ma}\gg\kappa_a,\kappa_m$. Then under either the blue-detuning driving (with a higher $\omega_p$ yielding negative $\Delta'$ and $\Delta$) and the red-detuning driving (with a lower $\omega_p$ yielding positive $\Delta'$ and $\Delta$), one can figure out $4$ situations for exacting a pair of nearly-resonant modes $m$ and $b$ or $a$ and $b$. We should stress that our STA approach is applicable to all of them. Now we choose the situation under the red-detuning driving and the higher-frequency hybrid mode is far-off-resonant from $b$, i.e., we have $\Delta'-\omega_b\approx 2g_{ma}/\sin2\phi\gg 0$.

Transforming the Hamiltonian (\ref{Hlinearnormal}) into the interaction picture with respect to $U=\exp[i\int_0^tds\Delta'(s)a^\dagger a+i\omega_b(m^\dagger m+b^\dagger b)t]$, we have
\begin{equation}\label{Hlinearrotat}
\begin{aligned}
H&=(\Delta-\omega_b) m^\dagger m \\&+g'\left\{a^\dagger be^{i[\int_0^tds\Delta'(s)-\omega_bt]}+a^\dagger b^\dagger e^{i[\int_0^tds\Delta'(s)+\omega_bt]}\right\}\\
&+g'^*\left\{ab^\dagger e^{-i[\int_0^tds\Delta'(s)-\omega_bt]}+abe^{-i[\int_0^tds\Delta'(s)+\omega_bt]}\right\}\\
&+g(m^\dagger b+m^\dagger b^\dagger e^{2i\omega_b t})+g^*(mb^\dagger+mbe^{-2i\omega_b t}).
\end{aligned}
\end{equation}
By discarding the fast-oscillating terms, the Hamiltonian turns out to be
\begin{equation}\label{Hlinearapp}
H=(\Delta-\omega_b)m^\dagger m+(g m^\dagger b+g^*mb^\dagger).
\end{equation}
That is exactly the effective Hamiltonian in Eq.~(\ref{Hlinear}) describing the lower-frequency hybrid mode coupled to the phonon mode.

\bibliographystyle{apsrevlong}
\bibliography{reference}
\end{document}